 \documentclass[useAMS,usenatbib, onecolumn]{mn2e}

\usepackage{epsfig}
\newcommand{\Nb}{{\bar N}}

\title[The amplitude of large-scale matter fluctuations] {Second-order matter fluctuations via higher order galaxy correlators}
\author[J. Bel and C. Marinoni]{J. Bel$^{1}$\thanks{E-mail:
jbel@cpt.univ-mrs.fr; marinoni@cpt.univ-mrs.fr} and C.
Marinoni$^{1,2}$\footnotemark[1]\\
$^1$ 
Centre de Physique Th\'eorique, Aix-Marseille Univ., CNRS UMR 6207, F-13288 Marseille cedex 9, France \\
$^2$ Intitut Universitaire de France, 103, bd. Saint-Michel, 75005 Paris, France} 

\begin{document}

\date{Accepted 2012 May 07; Received 2012 April 05; in original form 2012 January 18}

\pagerange{\pageref{firstpage}--\pageref{lastpage}} \pubyear{2002}

\maketitle

\label{firstpage}

\begin{abstract}

We provide a formula for extracting the value of the $rms$ of the linear  matter fluctuations on a scale $R$  directly from redshift survey data.
It allows us to constrain the real-space amplitude  of  $\sigma_R$ without requiring  any modeling of  the nature and  power spectrum of the matter distribution.
Furthermore, the formalism is completely insensitive to the character of the  bias function, namely its eventual scale or non-linear dependence. 
By contrasting measurements of $\sigma_R$  with predictions from  linear perturbation theory,  one can test for eventual  departures from 
the standard description of gravity on large cosmological scales.  

The proposed estimator  exploits the information contained in  the reduced  one-point and  two-point cumulant moments of the matter and galaxy density fields,   
and it can be applied on cosmic scales where linear and semi-linear perturbative approximations of the evolution of matter overdensities 
offer a satisfactory description of the full underlying theory.
We implement  the test with N-body simulations to quantify potential systematics and successfully  show that we are able to recover 
the present day value of $\sigma_8$ `hidden' in the simulations. We also design  a consistency check to gauge the soundness of the 
results inferred when the formalism is applied to real (as opposed to simulated) data.  We expect that  this approach will provide a sensitive probe of  
the clustering of matter when applied to  future large redshift survey such as  BigBOSS and EUCLID. 

\end{abstract}


\section{Introduction}

According to a widely accepted paradigm, cosmic structures grew from tiny dark matter density fluctuations 
present in the otherwise homogeneous and rapidly expanding early universe. 
The standard version of the model incorporates the assumption that these primordial and
Gaussian distributed fluctuations are amplified by gravity eventually turning into 
the rich structure that surrounds us  today.   This picture in which gravity, as described by general relativity, is
the engine driving cosmic growth is generally referred to as the
gravitational instability paradigm (GIP). However plausible it may seem, it is important to test its validity. 

In the local universe the  GIP paradigm has been shown to make sense of a vast amount of
independent observations on different spatial scales,  from galaxies to
superclusters of galaxies \citep[e.g.][]{pea01,teg06, reyes}.  Deep galaxy surveys 
now allow us to test whether the predictions of this
assumption are also valid at earlier epochs \citep{guzzo, blake}. In particular,  they allow us 
to asses  weather GR is the correct theory describing the action of gravity on large
cosmological scales \citep[e.g.][]{jza, uza09, aqga}.
 
Indeed, modifications to GR have been proposed
as alternatives to explain observations showing that the universe is undergoing accelerated expansion \citep{dvali, capozziello, ame,  buz08}.
Non standard gravitational models have also been invoked as an alternative to dark matter \citep{mil, bek} or to its 
standard physical characterization \citep{PiMa, ber08}.
Since these modified gravity theories are specifically tuned to explain the uniform cosmic  expansion history, a possible way to test 
their reliability is to analyze the inhomogeneous section of the universe, i.e.  cosmological perturbations of matter  \citep{lin05, zhang}.
 
It is a non-trivial task, compounded by our poor knowledge of the biasing mechanism, to trace  the global matter distribution 
by its luminous subcomponent.  An operational definition of bias, that is useful for investigating the hierarchical clustering of matter, 
is conventionally given in terms of continuous density fields, provided that the galaxy distribution  is smoothed on scales $R$  large enough compared to those 
where non-gravitational physics  operates. One can thus expand
the dimensionless density fluctuations of galaxies $\delta_{g,R}({\bf x})$ at position ${\bf x}$   in Taylor series of the underlying mass  
overdensity  $\delta_R$  at the same point

\begin{equation} 
\delta_{g,R}({\bf x})=\sum_{i=0}^{N}\frac{b_i}{i!}\delta_R^i({\bf x})
\label{biasfunction}
\end{equation}
\noindent where $b_i$ are the bias coefficients. It has been shown that,  in this large scale limit,  such a local transformation preserves the hierarchical properties of one-point  matter statistics
\citep{FG}.

The problem of interfacing theory (mass) with observations (light)  stems from the fact that the equations that give access to the value
of fundamental gravitational quantities (such as  the $rms$ of linear matter fluctuations on a given scale $R$ $\displaystyle (\sigma_R=\sqrt{\langle  \delta_R^2 \rangle})$ 
or the growth rate of linear perturbations $f=d \ln \delta/ \ln a$, where $a$ is the cosmic scale factor) are also the very same equations  that allows us to extract the value of  the bias parameters $b_i$. Since  the relevant physical and cosmological quantities are generally degenerate
with the bias parameters, it is not immediately obvious how  to fix their values. Because of this,  the traditional  approach consists in  assuming that gravitational and cosmological parameters are independently determined  to fix  the amplitude of the bias coefficients $b_i$ \citep[e.g.][]{lah02}.  

The viability of the  opposite route, that is investigating the coherence of the physical model given 
an {\it a-priori} knowledge of the bias function,  has been systematically explored only recently.  \cite{zhang} and \cite{sope}, for example,  have proposed to 
assess  the soundness of GR  by constructing  statistical indicators that are in principle insensitive to the linear biasing parameter, i.e.   
the lower order term in equation  (\ref{biasfunction}).  \citet{guzzo}, on the contrary,  tested GR by comparing the observed and predicted  growth rate  of matter fluctuations using data from deep redshift surveys. To fulfill their goal they adopted  the value of the linear biasing parameter provided  by independent observations,  such as the level of anisotropy in the Cosmic
Microwave Background \citep{koma} or the mean number density of galaxy clusters \citep{bor01,schu03}. Both these strategies suffer from the fact that there are now convincing  evidences about the non-linear character of the bias function \citep{mar05, gnbc, mar08, Kovac}. These testing scheme is also far from being economic, requiring data from multiple and independent probes of the large scale structure, redshift surveys, imaging surveys, CMB observations. 

An orthogonal, more general approach,  aims at extracting  from redshift surveys  both the value of $\sigma_R$  and the bias  parameters $b_i$. Several authors have shown that,  if the initial perturbations are Gaussian and if the shape of  third order statistics such as  the reduced skewness $S_3$ \citep{gazta94, gaztafrie94},  the bispectrum \citep{Fry94, scoc98, feld01, Verde} or the 3-point correlation function \citep{gnbc, gaztascoc, panszapudi}  is correctly described by results of the weakly non-inear perturbation theory, then one  can fix the amplitude of $b_i$ up to order 2 in a way that  is  independent from the overall amplitude of clustering (e.g. $\sigma_8$) and depends  only on the shape of the linear power spectrum.

In this paper we further explore the potentiality of this approach.  We exploit the information  encoded in a different kind of third order statistics, the reduced two-point correlator $C_{12}$  \citep{BernardeauC}. Following a suggestion of \citet{Szapudi98} we work out the explicit expressions for the bias coefficients up to order 4 as a function of $C_{12}$ and of  the reduced one-point cumulant $S_i$ (up to  order $5$).  Using these results we construct  the central formula of this paper (cfr. eq. (\ref{sigestimz})), i.e. an  estimator of  $\sigma_R$ that is independent from any assumption about  the linear power spectrum of matter.   As a result, by contrasting the redshift evolution of $\sigma_R$
with predictions of perturbation theory at linear order we can place constraints on the reliability of the GIP paradigm and in particular on the viability  of general relativity on large cosmological scales. Our  formalism  relies upon theoretical assumptions  and approximations that we have tested 
using   cosmological N-body simulations.  In this way we can  assess the overall coherence of the method  and  the impact  of  potential systematics.

The paper is organized as follows: in \S 2, we introduce the relevant definitions and relations  that provide the standard framework for understanding  the clustering of matter in the weakly 
non-linear regime. We work out  the relationships between the two-point reduced correlators of  galaxies and matter in \S 3, and in \S4 we show how to use them in order to express  the bias coefficients $b_i$ (up to order 4) as a function of observable quantities. The central achievement of this paper is presented in \S 5  where we detail  the computational scheme that allows  to extract both the amplitude and redshift scaling of $\sigma_R$ in real space.  In section \S 6 we work out the formalism for  mapping continuum variables into discrete observables and we show how to implement in practice the proposed testing strategy using  simulations. We also  design a strategy to test the coherence of the results. Conclusions and prospects are drawn in \S 7.

\section{Higher order correlations as a measure of the large-scale distribution of matter}

In this section we give the theoretical background for analyzing the clustering of matter on large cosmic scales as well as the mathematical framework  
to the model presented in Section \S3.  We first introduce our notation and  briefly overview the current understanding of the hierarchical clustering of matter in the universe.
We then discuss the high order statistical descriptors of the density perturbation field and display  some fundamental results
about their linear and weakly non-linear  evolution. We finally show how these results are changed when the relevant statistics are 
sampled with a given  finite resolution. A thorough treatment of the subject can be found in the review of \citet{revb}.

\subsection{Joint $K$-point cumulant moments of the  continuous fluctuation field}
Be  $\lambda({\bf x})$  the intensity function describing the overall mass distribution in the universe, i.e. a stochastic field representing the density of  matter at 
a given position ${\bf x}$ \citep[e.g.][]{masa}.  We consider $\lambda({\bf x})$ as a particular (homogeneous and isotropic) realization drawn from an ensemble $\mathcal{E}$ and  
indicate with $\mathcal{F}\big[ \lambda ({\bf x}) \big]$ its probability density functional (PDF). 

Since the intensity field is positively defined,  $\mathcal{F}\big[ \lambda ({\bf x}) \big]$ is  by definition  non Gaussian. Its complete characterization requires the knowledge 
of  the entire (formally infinite)  hierarchy of  the  $K$-point expectation values  

\begin{equation}
\big \langle \lambda({\bf  x}_1)\, ... \,\lambda({\bf x}_K) \big \rangle = i^{-K} \frac{ {\displaystyle \delta^{K}  \mathcal{M}_{\lambda}[J]}}{{\displaystyle \delta J({\bf x}_1) \, ... \,  \delta J({\bf x}_K)}}  \Bigr |_{J({\bf x}_1)=...=J({\bf x}_K)=0}
\label{funcder}
\end{equation}

\noindent where
 
\[
\mathcal{M_{\lambda}}[J]= 
\int   \mathcal{D}  \lambda({\bf x})  \mathcal{F} \big[ \lambda({\bf x})\big] e^{ i \int d{\bf x} J({\bf x})\lambda({\bf x})}= \Big \langle  e^{ i \int d{\bf x} J({\bf x})\lambda({\bf x})} \Big \rangle
\]

\noindent is the $K$-point moment generating functional,  and where $\mathcal{D}  \lambda({\bf x}) $ represents a suitable measure introduced in $\mathcal{E}$ 
such that the total probability turns out to be normalized to 1 \citep{mlb}.

The cosmological principle guarantees the existence of a non-zero  value for  the expected value of the $\lambda$ field. It is thus convenient  to characterize
inhomogeneities in the matter distribution in terms of the  local dimensionless density contrast  

\begin{equation}
\delta( {\bf x} ) \equiv \frac{\lambda({\bf x} )}{\langle\lambda({\bf x})\rangle}-1.
\end{equation}

\noindent We assume  that the ensemble average over $\mathcal{E}$  is equivalent to averaging a particular realization over different spatial positions. 
If this operation is done on a fair sample then $\langle \delta ({\bf x}) \rangle=0$. 

An useful tool for probing the large-scale cosmological structure, because of their relatively simple connection to both theory and observations,  are the
joint $K$-point cumulant moments of order ${\bf n}=(n_1, n_2.....n_K)$ of the cosmic overdensity field
\begin{equation}
 \kappa_{n_1, ..., n_K}({\bf x}_1, ..., {\bf x}_K) \equiv  \big \langle \delta^{n_1}({\bf x}_1) \, ... \, \delta^{n_K}({\bf  x}_K) \big \rangle_c,
\label{kappa}
\end{equation}

\noindent and, in particular,    the irreducible $K$-point autocorrelation functions

\begin{equation}
 \kappa_{1, ..., 1}({\bf x}_1,..., {\bf x}_K) \equiv  \big \langle\delta({\bf x}_1) \, ... \, \delta({\bf x}_K) \big \rangle_c
\end{equation}

\noindent  generally   denoted as $\xi_K$ and shortly  called correlation functions.   
By definition, the  generating functional of the $K$-point  cumulant moment  of the overdensity field 
is the logarithm of the moment  generating  functional 

\begin{equation}
{\displaystyle \mathcal{C}[J] \equiv \ln  \mathcal{M}_{\delta}[J].}
\end{equation}

The advantage of  computing connected averages  instead of statistical averages 
($\mu_{n_1, ..., n_K} \equiv  \big \langle \delta^{n_1}({\bf x}_1) \, ... \, \delta^{n_K}({\bf  x}_K) \big \rangle$)  is that  cumulants 
are zero if the random variables representing the 
value of the stochastic field at different spatial positions are statistically
independent. Conversely,  a cumulant is not zero if and only if  the random variables in it  are  statistically ``connected". 
A specific consequence of this properties is that  $\kappa_{n_1, ..., n_K} \rightarrow 0$ as any subset of  positions ${\bf x}_i$ are displaced  to infinite separation. 
Finally, cumulants  can be explicitly represented in terms of only the lower order moments, that is as a function of  $\mu_{m_1...m_K}({\bf x }_1,...,{\bf x}_K)$ with 
$0 \leq m_i \leq n_i$.   This follows  from  taking successive functional derivatives of the generating functional $\mathcal{C}[J]$ 
\begin{equation}
\kappa_{n_1, ..., n_K}= i^{-N} \frac{{\displaystyle  \delta^{N}  \mathcal{C}[J]}}{{\displaystyle \delta J^{n_1}({\bf x}_1) \, ... \,  \delta J^{n_K}({\bf x}_k)}} \Bigr |_{ J({\bf x}_1)=...=J({\bf x}_K)=0}
\label{funcder2}
\end{equation}

\noindent where $N=\sum_{i=1}^{K} n_i$. The result can be formally written as   \citep{meeron}

\begin{equation}
\kappa_{n_1, ..., n_K} =   {\displaystyle - \prod_{j} \nu_{j}!  \sum_{l=1}^{N}  \sum_{\gamma_i, m_{ij}}}   \big(\sum  \gamma_i-1\big )!  \big (-1\big )^{ \sum \gamma_i}
  {\displaystyle \prod_{i=1}^{l} \frac{1}{\gamma_i!}} \Bigg\{  \frac{ \big \langle \prod_{j=1}^{K} \delta^{n_{ij}}  ({\bf x_j})  \big \rangle} {\prod_{j}  n_{ij}!}  \Bigg\}^{\gamma_i}.
\label{epca}
\end{equation}

\noindent In this last equation,  $\gamma_i$ and $m_{ij}$ are non-negative integers satisfying the set of equations

\[ \sum_{i=1}^{l}  \gamma_i m_{ij}=n_j \]

\noindent and bookkeeping all the possible decompositions

\[ \big(\delta^{n_1}({\bf x}_1) \, ... \, \delta^{n_K}({\bf x}_K)\big)\rightarrow \prod_{i=1}^{l}   \big(\delta^{m_{i1}}({\bf x}_1) \, ... \,  \delta^{m_{iK}}({\bf x}_K)\big)^{\gamma_i}.\]

In this study we are interested in the joint cumulant moments  taken at two different locations ${\bf x}_1$  and ${\bf x}_2$  up to the order  
$N=n_1+n_2=7$. For these statistics, from now on simply called  correlators and indicated as $\kappa_{nm}$,  eq. \ref{epca} gives 

\begin{eqnarray}
\kappa_{{11}} & = &  \mu_{{11}}  \label{expa} \\
\kappa_{{12}} & = &  \mu_{{12}}  \nonumber \\ 
\kappa_{{13}} & = &\mu_{{13}}-3\,\mu_{{11}}\mu_{{2}}   \nonumber \\
\kappa_{{22}} & = & \mu_{{22}}-2\,{\mu_{{11}}}^{2}-{\mu_{{2}}}^{2} \nonumber \\
\kappa_{{14}} & = & \mu_{{14}}-6\,\mu_{{12}}\mu_{{2}}-4\,\mu_{{11}}\mu_{{3}} \nonumber \\
\kappa_{{23}} & = & \mu_{{23}}-6\,\mu_{{12}}\mu_{{11}}-\mu_{{3}}\mu_{{2}}-3\,\mu_{{12}}\mu_{{2}} \nonumber \\
\kappa_{{15}} & = & \mu_{{15}}-5\,\mu_{{11}}\mu_{{4}}-10\,\mu_{{12}}\mu_{{3}}-10\,\mu_{{13}}\mu_{{2}}+30\,\mu_{{11}}{\mu_{{2}}}^{2} \nonumber \\
\kappa_{{24}} & = & \mu_{{24}}-4\,\mu_{{12}}\mu_{{3}}-8\,\mu_{{13}}\mu_{{11}}-6\,\mu_{{22}}\mu_{{2}}-6\,{\mu_{{12}}}^{2}+
    6\,{\mu_{{2}}}^{3}+24\,{\mu_{{11}}}^{2}\mu_{{2}}-\mu_{{4}}\mu_{{2}} \nonumber  \\
\kappa_{{33}}& = & \mu_{{33}}-6\,\mu_{{13}}\mu_{{2}}-{\mu_{{3}}}^{2}-9\,{\mu_{{12}}}^{2}+12\,{\mu_{{11}}}^{3} -9\,\mu_{{22}}\mu_{{11}}+
         18\,\mu_{{11}}{\mu_{{2}}}^{2} \nonumber \\
\kappa_{{16}}& = &  \mu_{{16}}-6\,\mu_{{11}}\mu_{{5}}-20\,\mu_{{13}}\mu_{{3}}-15\,\mu_{{12}}\mu_{{4}}-15\,\mu_{{14}}\mu_{{2}}+
         120\,\mu_{{11}}\mu_{{2}}\mu_{{3}}+90\,\mu_{{12}}{\mu_{{2}}}^{2} \nonumber \\
\kappa_{{25}}& = &   \mu_{{25}}+40\,{\mu_{{11}}}^{2}\mu_{{3}}-10\,\mu_{{14}}\mu_{{11}}-5\,\mu_{{12}}\mu_{{4}}-20\,\mu_{{13}}\mu_{{12}}+
          30\,\mu_{{12}}{\mu_{{2}}}^{2} -10\,\mu_{{22}}\mu_{{3}}-10\,\mu_{{23}}\mu_{{2}}-\mu_{{5}}\mu_{{2}}+
          120\,\mu_{{3}}{\mu_{{2}}}^{2} \nonumber \\ 
          & & +20\,\mu_{{11}}\mu_{{2}}\mu_{{12}} \nonumber \\
\kappa_{{34}}& = &  \mu_{{34}}-4\,\mu_{{13}}\mu_{{3}}-12\,\mu_{{13}}\mu_{{12}}-18\,\mu_{{22}}\mu_{{12}}-3\,\mu_{{14}}\mu_{{2}}+
          24\,\mu_{{11}}\mu_{{2}}\mu_{{3}}+36\,\mu_{{12}}{\mu_{{2}}}^{2}-6\,\mu_{{23}}\mu_{{2}}+6\,\mu_{{3}}{\mu_{{2}}}^{2}+ \nonumber \\
          & &  72\,{\mu_{{11}}}^{2}\mu_{{12}} -12\,\mu_{{23}}\mu_{{11}}-\mu_{{3}}\mu_{{4}}+72\,\mu_{{11}}\mu_{{2}}\mu_{{12}} \nonumber \\
          \nonumber
\end{eqnarray}

\noindent where we have used the fact that $\langle \delta \rangle=0$. In  the one-point limiting case 
one recovers the expressions of the cumulant moments $\kappa_{N}$  given by 
\citet{Fry84} (cfr. eq.  17). Note, also,  that correlators are symmetric  with respect to exchanging the indexes.

\subsection{Dynamics of the mass fluctuations}

Cosmological  perturbation theory
can be successfully applied to  determine the evolution of the hierarchy of moments and correlators of the mass
fluctuations.  Theoretical predictions, however,  cannot be derived without the prior assumption of a specific, primordial,  clustering model.

An interesting, simple and physically motivated case is that in which the initial density fluctuations in the matter component
are described by a Gaussian probability density functional. In this case, odd moments are zero and all the even moments are completely specified by the moment of order two.
However, even starting with an initial  Gaussian over-density field,  gravitational dynamics induces non-zero higher order correlations already 
in the mildly non-linear regime. In particular, by applying weakly non-linear perturbation theory (WNLPT) to compute the correlations that first emerges 
from a gravitationally unstable  Gaussian field, one finds  that,  at leading order (lowest order in $\kappa_{2}$), 

\begin{equation}
\displaystyle  \kappa_{N} = S_N  \kappa_{2}^{N-1} 
\label{hs1}
\end{equation}

\noindent where $S_N$ are structure constants that do not depend on the scale of the fluctuations, 
on the initial power spectrum and are almost insensitive to cosmological parameters  \citep{pee80, Fry84, bernardeau92}.

While extremely  useful to characterize the gravitational  properties of matter, the one-point reduced moments $S_N$ do not retain  information about the spatial 
configuration of clustering.  Interestingly, their most immediate generalization, i.e.  the  correlators $\kappa_{n m}$  show 
hierarchical scaling properties similar to those in eq. \ref{hs1}. Indeed,  by assuming  that  the locations 
${\bf x_1}$ and ${\bf x_2}$ are sufficiently separated (large separation (LS) limit), 
\citet{BernardeauC} showed that 

\begin{equation}
\kappa_{nm}=C_{nm}\kappa_{11}  \kappa_{2}^{n+m-2}.
\label{hs2}
\end{equation}

\noindent As it is the case for the $S_N$, even the $C_{nm}$ statistics (hereafter calledthe reduced correlators of order $(n,m)$) do show  a negligible dependence on 
the cosmological mass density parameter, the cosmological constant and
cosmic epoch \citep{BernardeauC}.   Indeed they are mostly sensitive to the initial conditions (assumed to be Gaussian)
and to the physical mechanism  that  drives the distribution of the  fluctuations away from the initial state (assumed to be gravity).

\noindent  In the LS approximation,  the reduced correlators $C_{nm}$ factorize as 

\begin{equation}
 C_{nm}=C_{n1}C_{m1}.
 \label{facto}
\end{equation}

\noindent This key property  shows that not all the reduced correlators are independent, and that some of them do not give access to 
complementary cosmological information.
 \subsection{Finite resolution effects on the dynamics of the field}

In order to ease the comparison with an intrinsically discrete process such as the distribution of galaxies, as well as to facilitate the incorporation of  the biasing scheme 
(e.g. eq. \ref{biasfunction}) into the analysis,  it is useful to smooth the mass distribution  on a spatial scale $R$. This is done  by convolving  the 
overdensity field $\delta$ with a (normalized) window function of size $R$.

\begin{equation}
{\displaystyle \delta_R({\bf x})=\int \delta({\bf x'}) W  \bigg [ \frac{\mid {\bf x}-{\bf x'} \mid}{R}  \bigg ]d^3{\bf x'}}.
\end{equation}

For a top-hat filter, $\delta_R({\bf x})$ is just the volume average of the density contrast over a sphere of radius $R$.  Note that 
the smoothed  correlators of order  $N=(n,m)$  retain some of the information contained in the $Nth$ order correlation function.
As a matter of fact,  
\begin{equation}
\kappa_{nm,R}= \frac{1}{V_R^{n+m}} \int_{V_R({\bf x}_1)} d{\bf y}_1 ...  d{\bf y}_n  \int_{V_R({\bf x}_2)} d{\bf y}_{n+1} ... d{\bf y}_{n+m}  \xi_{n+m}({\bf y}_1, ..., {\bf y}_{n+m})
\end{equation} 

\noindent where we have assumed   a top-hat filter of volume $V_R$. From a physical point of view $\kappa_{nm,R}$ is the average of the correlation
function of order $n+m$ over two distinct volumes separated by  $|{\bf x}_1-{\bf x}_2|$. 

Smoothing and averaging are non commutative operations. As a consequence, while the relations given in eqs. \ref{expa},  \ref{hs1} and \ref{hs2} retain their validity
when applied to filtered fields,  the amplitudes of the  smoothed dynamical variables (cumulants and reduced cumulants) become, instead,  scale dependent. 
Consider for example  the lowest-order  non-zero cumulant moment and correlator, that is the variance of the mass fluctuations on a scale R

\begin{equation}
\sigma^2_R=\kappa_{2,R} = \big \langle \delta_R^2({\bf x}) \big \rangle_c, 
\end{equation}
\noindent  and the covariance of the  smoothed mass overdensity field 

\begin{equation}
\xi_R(r) = \kappa_{11,R}= \big \langle \delta_R({\bf x}) \delta_R({\bf x} +{\bf r}) \big \rangle_c. 
\end{equation}

\noindent Suppose, further,  that mass fluctuations are small  ($|\delta| \ll 1 $) and described by the linear (dimensionless) power spectrum

\begin{equation}
\Delta^2_L = 4\pi A k^{n_s+3}T^2(k), 
\label{pws}
\end{equation}

\noindent where   $A$ is the normalization factor, $n_s$ the primordial spectral index,  
and $T^2(k)$ the transfer function \citep{BBKS,EBW, EH}.  This implies that  the amplitudes of second order statistics  
evolve as a function of time and scale as 

\begin{equation}
\sigma^2_R(z)=\sigma^2_8(0)D^2(z){\mathcal F}_R,
\label{teosigma}
\end{equation}

\noindent and

\begin{equation}
\xi_R(r,z)=\sigma_8(0)^2 D^2 (z){\mathcal G}_R(r).
\label{csir}
\end{equation}

\noindent The normalization of these equations is conventionally fixed at a scale  $r_8=8h^{-1}$Mpc,  $D(t)$ represents  the linear growing mode \citep{pee80},  while the effects of filtering are incorporated in the functions

\begin{equation}
{\mathcal F}_R=\frac{\int_0^{+\infty}\Delta_L^2(k)\hat{W}^2(kR)d\ln k}{\int_0^{+\infty}\Delta_L^2(k)\hat{W}^2(kr_8)^2d\ln k}
\label{ff}
\end{equation}
and
\begin{equation}
{\mathcal G}_R(r)=\frac{\int_0^{+\infty}\Delta_L^2(k) \hat{W}^2(kR)\frac{\displaystyle \sin(kr)}{\displaystyle kr}d\ln k}{\int_0^{+\infty}\Delta_L^2(k)\hat{W}^2(kr_8)d\ln k}
\label{gg}
\end{equation}
\noindent where $\hat{W}$ is the Fourier transform of the window function.  In Figure 1 we show the scaling of the smoothed two-point correlation function (as a function of 
both $R$ and $r$)  at two different cosmic epochs.  There is an overall qualitative resemblance between  the $r$ dependence of the two-point correlation function $\xi(r)$ and the  $R$ dependence of its smoothed version $\xi_R(r)$. More interestingly,   the characteristic non-monotonic scaling induced by the baryon acoustic oscillations (BAO) that are  frozen in the large scale matter distribution survives to the smoothing procedure and stands out  also in the second order correlator as soon as $r$ approaches $\sim 100h^{-1}$Mpc.

Computing the amplitude of  the smoothed reduced cumulants and correlators  at  the next order (i.e. $S_{3,R} $ and $C_{12,R}$) requires results from the weakly non-linear perturbation theory. 
If the primordial mass field is Gaussian and  fluctuations with  wavelength $<<R$ are suppressed using  a top-hat filter,  then 
the third order  reduced moment, which is often referred to as the skewness of the density field,  is \citep{juskie, BernardeauB} 

\begin{equation}
S_{3,R}=\frac{34}{7} + \gamma_R
\label{s3prediction}
\end{equation}

\noindent while, in the LS  limit, that is for  separations $r>>R$,  the reduced correlator of the same order 
is   \citep{BernardeauC}

\begin{equation}
C_{12,R}(r)=\frac{68}{21} + \frac{1}{3}\gamma_R+ \frac{1}{3}\beta_R(r).
\label{c12prediction}
\end{equation}

\noindent The  effect of filtering is to introduce additional,  scale-dependent,  coefficients

\begin{equation} 
\gamma_R \equiv\frac{d \log\sigma^2_R}{d\log R}
\label{gamma}
\end{equation}
\begin{equation}
\beta_R(r)\equiv\frac{d \log\xi_R(r)}{d\log  R},
\label{beta}
\end{equation}

\noindent such that  the reduced moment $S_{3,R}$ effectively  depends on the local slope of the linear power spectrum of density fluctuations 
(decreasing with the slope of the power spectrum), while the reduced correlator  $C_{12,R}$  acquires a specific  and characteristic non-local dependence. 
In the following we parameterize the distance between the centers of independent smoothing spheres as $r=nR$, where $n$ is a
generic real parameter (usually taken, without loss of generality, to be an integer).  According to the analysis of \citet{BernardeauC}, the LS
regime is fairly well recovered as soon as  $n \ge 3$. Interestingly, since WNLPT  results hold for large $R$,  such a small $n$ defines 
a separation scale $r$ that is already accessible using current redshift surveys  such as the Sloan Digital Sky Survey.

The  $\beta_R$ contribution in eq. (\ref{beta})   is usually neglected  \citep{BernardeauC}
since, in the LS limit,   $\xi_R(r) $ is simply the two-point correlation function of the un-filtered field, a function
that effectively vanishes for large separations.  This is a critical simplification and the domain of its validity deserves more in-depth analysis.
By assuming a power-law spectrum of effective index $n_{e}=-1.2$ (i.e. $\gamma_R=-(n_e+3)=-1.8)$,  we obtain that, on all scales $R$, the amplitude of $\beta_R(nR)$ 
becomes negligible ($ \le 0.08$) as soon as $n \ge 3$. Anyway this rapid convergence to zero is a peculiar characteristic of a scale-free power spectrum. 
If we  consider a more realistic  power spectrum (cfr. eq. \ref{pws})
on scales  that are accessible to both semi-linear theory and current large scale data (i.e. $10<R<30h^{-1}$ Mpc, and $n \sim 3$),  the amplitude of the $\beta_R$ contribution is still  significant and varies non monotonically  as a function of  the length scales $r$ on which the cell correlation is estimated. This is  illustrated  in  Figure \ref{betatheo} where  we contrast the scaling of  $\beta_R(nR)$ and $\gamma_R$ for different vales of $R$ and $n$. 
The systematic error in the  estimation of $C_{12}$  that is  induced by neglecting the $\beta_R$-term on relevant cosmological scales,  is larger than the error with which this statistics can already be estimated  from current data (see Figure \ref{d3bigboss} in section \S 6.2).  For example, 
 the amplitude of  $\beta_R$, for characteristic values  $n=3$ and $R= 10(/25) h^{-1}$Mpc, is 
$\sim 15(/30)$ per cent that of $\gamma_R$.  Interestingly,  one can see that, as for $\gamma_R$,  also  the value of $\beta_R(r)$ does not depend on cosmic time,  at least at linear order. 
This redshift-independence follows immediately from eqs. (\ref{csir}) and (\ref{beta}).

Notice, finally, that WNLPT theory results are expected to hold in the correlation length range  in which  $C_{12}$ can be  unambiguously defined, that is  up to 
the scale where the correlator $k_{11}$ crosses zero (see eq. \ref{hs2}). This requirement sets an upper limit to the effective  correlation scale $n$ that can be investigated
using predictions of WNLPT.  In this large scale context, also notice that non-linear effects  contributing to  the baryon acoustic peak in the 
the two-point correlation function, would modify  predictions obtained on the basis of the simple linear model of eq. (\ref{pws}). For these reasons we limit the present analysis 
to correlation scales  $nR \leq 100h^{-1}$ Mpc.  In a future work (Bel et al. in prep) we will compare WNLPT predictions against fully non-linear numerical results from matter simulations 
in order to  constrain in a quantitative way the boundaries of the interval where WNLPT results safely apply.

\begin{figure}
\begin{center}
\includegraphics[width=8.5cm]{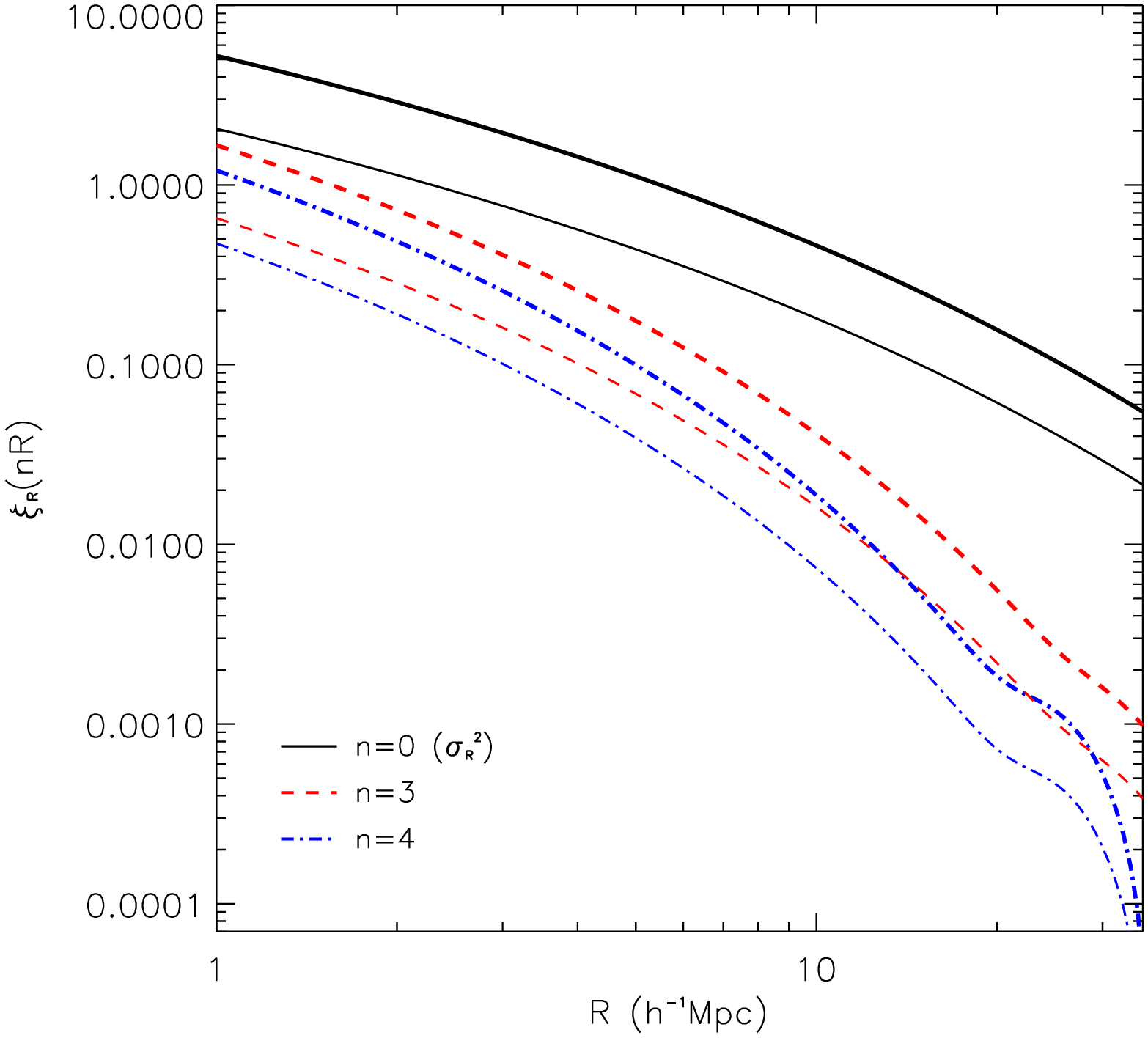} 
\includegraphics[width=8.5cm]{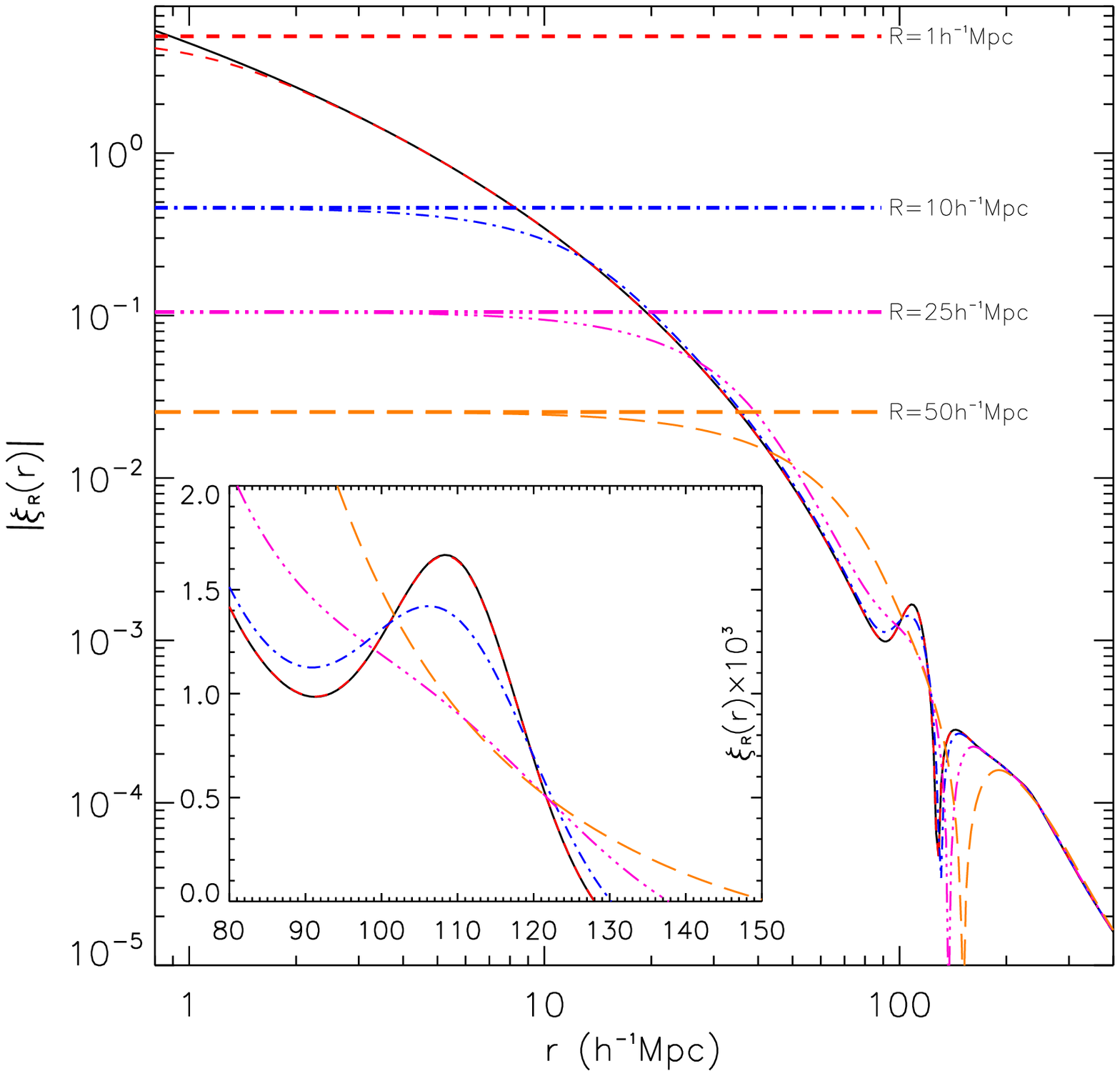} 
\end{center}
\caption{\small {\it Left: }  the $2$-point correlation function of the smoothed mass density field $\xi_R$  is shown as a function of the smoothing scale $R$.
We plot $\xi_R$   for different values of the correlation length $r=nR$ i.e. $n=0$ (black solid lines), $n=3$ (red dashed lines), and $n=4$ (blue dot-dashed lines) and at two different 
redshifts: $z=0$ (thick lines) and $z=1$ (thin lines). We adopt the linear power spectrum model of \citet{EH} and we assume the  following set of cosmological parameters: $\Omega_m=0.26$, $\Omega_\Lambda=0.74$, $H_0=72$ km s$^{-1}$ Mpc$^{-1}$, $\sigma_8(0)=0.79$,    $\Omega_b=0.044$ and the spectral index $n_s=0.96$. Note that, for $n=0$,  $\xi_R(0)=\sigma^2_R$, while for a correlation length that goes to infinity the correlation function of the smoothed density field  tends to zero. The characteristic  bump induced by the baryon acoustic oscillations becomes clearly visible as soon as $n$ is large enough.
 {\it Right:}  the $2$-point correlation function $\xi_R(r)$  is shown as a function of the correlation length $r$ for fields smoothed on  different scales, that is $R=1, 10, 25$ and $50 h^{-1}$Mpc. 
As the smoothing scale $R$ tends to zero, one recovers the $2$-point correlation function of matter particles, while as the filtering scale becomes larger,  the BAO peak is progressively suppressed.
Note also that for $r \rightarrow 0 $, the value of $\xi_R(r)$ saturates to the variance of the field on the given scale $R$. On the opposite sense ($r \rightarrow \infty$) all the curves converge to the value of the two-point  correlation function of matter particles.}
\label{xitheo}
\end{figure}

\begin{figure}
\begin{center}$
\begin{array}{cc}
\includegraphics[width=8.5cm]{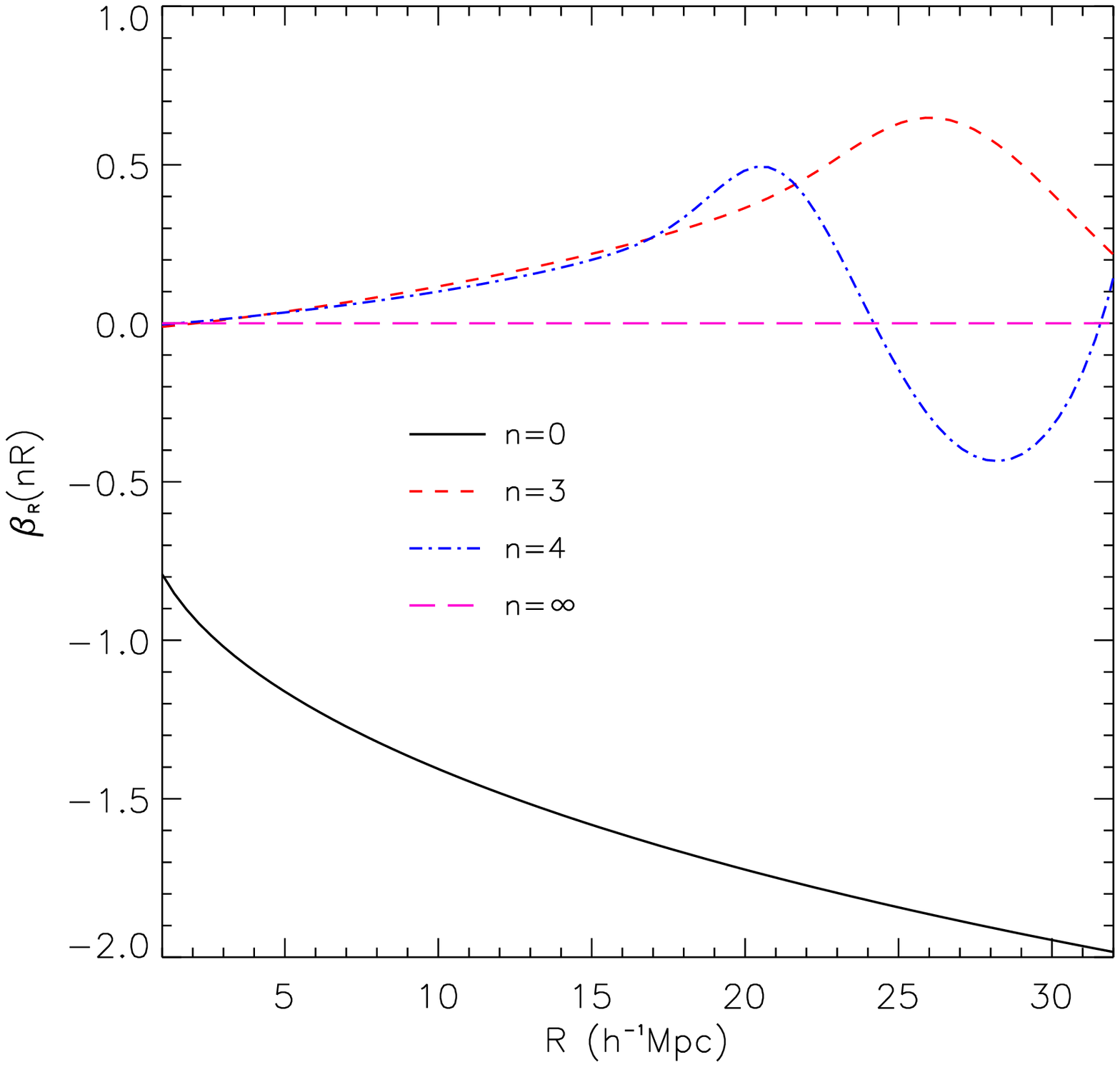} &
\includegraphics[width=8.5cm]{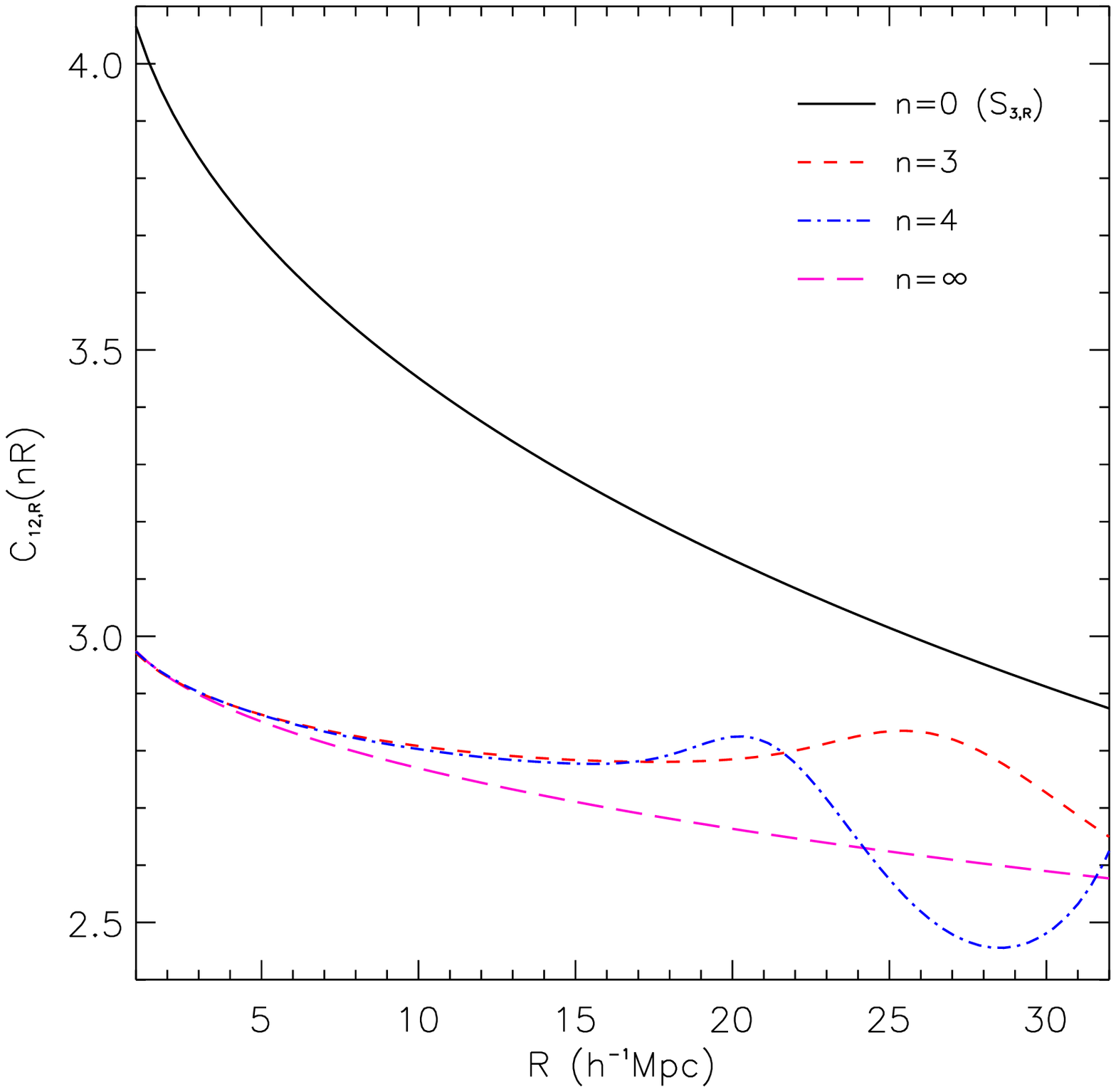} 
\end{array}$
\end{center}
\caption{\small   {\it Left :}  the logarithmic derivative of $\xi_R(r)$ with respect to $R$ (cfr. eq. \ref{beta}) is shown as a function of the smoothing scale $R$ for different values of the correlation length $r=nR$ (i.e. for $n=0$ solid line, $n=3$ short-dashed line, $n=4$ dot-dashed line and $n=\infty$ long-dashed line).  We have assumed the   linear power spectrum model of \citet{EH} and the  same parameters value listed in the caption of Figure 1.
The non-monotonic scaling  of $\beta_R(nR)$ at a scale R$\sim 25 h^{-1}$Mpc  is induced by the baryon acoustic oscillations. Note that, in the linear regime,  $\beta_R(r)$ does not depend on the cosmic epoch, i.e. it is a redshift independent quantity.
{\it Right :} the scaling of the reduced correlator of order $3$ is shown as a function of  the smoothing scale $R$ for  different values of the correlation length $r=nR$.
Note that for $n=0$ the correlator reduces to the skewness $S_{3,R}$ and  that for $n=\infty$ the expression of $C_{12,R}(nR)$ reduces to the one adopted by \citet{BernardeauC}.  }
\label{betatheo}
\end{figure}

\section{Galaxy Correlators}

Smoothing  is not the only process that leaves an imprint on the value of cumulant moments.
Their  amplitude and shape  is also altered if galaxies, instead of massive particles, are used to trace the overall matter distribution.
\citet{FG}  computed the  effects of biasing  on one-point (smoothed) cumulant moments up to order $5$. In this section, we generalize those results by deriving the expressions of the 
smoothed two-point cumulant moments of the galaxy overdensity field up to  the same order. 
These are new observables that can be used to test predictions of the GIP  in the weakly non-linear regime, to give  insight into the gravity induced   
large-scale bias, and also  to distinguish models with Gaussian initial conditions from their non-Gaussian alternatives.  Some of these possibilities will be explored in sections \S 4 and \S 5. Since from now on we  only consider smoothed statistics, we simplify our notation by omitting any explicit reference to  the smoothing scale $R$. 
The $R-$dependence  of relevant statistical quantities will be  re-emphasized when necessary.

Consider the  $N$-point, smoothed,  correlation functions of matter up to order $5$

\begin{eqnarray}
\xi_{ij} & = & \kappa_{11}({\bf x}_{i}, {\bf x}_{j}) \label{defi}   \\
\zeta_{ijk} & = & \kappa_{111} ({\bf x}_{i}, {\bf x}_{j}, {\bf x}_{k}) \nonumber  \\
\eta_{ijkl} & = & \kappa_{1111} ({\bf x}_{i}, {\bf x}_{j}, {\bf x}_{k}, {\bf x}_{l})  \nonumber \\
\omega_{ijklm} & = & \kappa_{11111} ({\bf x}_{i}, {\bf x}_{j}, {\bf x}_{k}, {\bf x}_{l}, {\bf x}_{m}). \nonumber  \\
\nonumber
\end{eqnarray}

\noindent At leading order, the corresponding  statistics describing the distribution of galaxies, labeled with the
suffix `$g$' are

\begin{eqnarray}
\xi_{12,g}    & =  & {b_{1}}^{2} \, \xi_{12}  \label{defibias} \\
\zeta_{123,g}   &  = & {b_{{1}}}^{3}\,  \zeta_{{123}}+ {b_{{1}}}^{3}c_{{2}} \big ( \xi_{{13}}\xi_{{23}}+2\ {\mathrm perm} \big )  \nonumber  \\ 
\eta_{1234,g}  & =  & {b_{{1}}}^{4} \, \eta_{{1234}}+{b_{1}}^{4} \big [ c_{2} ( \xi_{{23}}\zeta_{{124}}+11\ {\mathrm perm})  +  c_{3} ( \xi_{{14}}\xi_{{24}}\xi_{{34}}+3\ {\mathrm perm})+ {c_{2}}^{2} (\xi_{{13}}\xi_{{24}}\xi_{{34}}+11\ {\mathrm perm} )\big ]  \nonumber \\
\omega_{12345,g}  & = & {b_{{1}}}^{5}\, \omega_{{12345}}+{b_{1}}^{5} \big [  c_{2} \left(\xi_{{15}}\eta_{{2345}}+19\ {\mathrm perm}+ 
 \zeta_{{125}} \zeta_{{345}}+14\ {\mathrm perm} \right)+ c_{3} \left(\zeta_{{135}}\xi_{{25}}\xi_{{45}}+29\ {\mathrm perm}\right)+  \nonumber \\
 & &  c_{4}\left(\xi_{{14}}\xi_{{24}}\xi_{{34}}\xi_{{45}}+4\ {\mathrm perm}\right)+ {c_{2}}^{2}\left(\zeta_{{135}}\xi_{{24}}\xi_{{25}}+119\ {\mathrm perm}\right)+
 c_{2}c_{3}\left(\xi_{{14}}\xi_{{25}}\xi_{{34}}\xi_{{24}} +59\ {\mathrm perm}\right) \nonumber \\
 & & + {c_{2}}^{3}\left(\xi_{{13}}\xi_{{24}}\xi_{{45}}\xi_{{15}}+59\ {\mathrm perm}\right) \big ]. \nonumber \\
 \nonumber
\end {eqnarray}

\noindent 
The mapping between correlators of mass ($\kappa_{nm}$)  and  galaxy ($\kappa_{nm,g}$) follows immediately  by taking the two-point limiting case in the above expressions.
Listed below are the  results up to  order $n+m=4$ (including also non-leading terms):

\begin{eqnarray}
\kappa_{11,g} & = & b_{{1}}^{2} \kappa_{11}+b_{{1}}^{2}  \big ( c_{{3}}+C_{{12}} c_{{2}} \big ) \kappa_{11} \, \kappa_{{2}}+ 1/2 \, b_{{1}}^{2} c_{{2}}^{2}  \kappa_{11}^{2} \label {hiera}  \\
\kappa_{12,g} & = &  b_{{1}}^{3} ( C_{{12}}+2\,c_{{2}}  ) \kappa_{11}\,\kappa_{{2}}+b_1^3c_2 \kappa_{11}^2+ b_{{1}}^{3} \big ( 5/2\,c_{{3}}C_{{12}}+2\, c_{{2}}^{2}C_{{12}}
  +3\,c_{{3}}c_{{2}}+c_{{4}}  +1/2 \,c_{{2}}C_{{22}}+ c_{{2}}^{2}S_{{3}}+c_{{3}}S_{{3}} \nonumber \\
  & & + c_{{2}} C_{{13}} \big )\kappa_{11}\, \kappa_{{2}}^{2}+ b_{{1}}^{3} \big ( c_{{2}}^{3}+ 1/2 \,c_{{4}}+3\, c_{{2}}^{2}C_{{12}} +  2\,c_{{3}}c_{{2}}+c_{{3}}C_{{12}} \big) \kappa_{11}^{2} \kappa_{{2}}+ b_{{1}}^{3}c_{{3}}c_{{2}} \kappa_{11}^{3}  \nonumber \\
   \kappa_{13,g} & =& {b_{{1}}}^{4}\big ( 6\,{c_{{2}}}^{2}+C_{{13}}+3\,c_{{2}}S_{{3}}+3\,c_{{3}}+6\,c_{{2}}C_{{12}} \big ) \kappa_{11}\,{\kappa_{{2}}}^{2}+{b_{{1}}}^{4} 
\big ( 3\,C_{{12}}c_{{2}}+6\,{c_{{2}}}^{2} \big ) {\kappa_{11}}^{2}\kappa_{{2}}+{b_{{1}}}^{4}c_{{3}}{\kappa_{11}}^{3}+  {b_{{1}}}^{4} \big ( 3/2\,c_{{2}}C_{{14}}\nonumber \\
& & +9/2\,C_{{12}}c_{{3}}S_{{3}}+12\,{c_{{2}}}^{2}c_{{3}}+6\,{c_{{2}}}^{3}S_{{3}}+45/2\,c_{{2}}c_{{3}}C_{{12}}+9\,c_{{2}}c_{{4}}+3/2\,c_{{3}}S_{{4}}+ 9/2\,c_{{4}}S_{{3}}+3/2\,c_{{5}}+15/2\,{c_{{3}}}^{2}+3\,{c_{{2}}}^{2}S_{{4}} \nonumber \\
 & & +3\,{c_{{2}}}^{2} C_{{22}} +15/2\,{c_{{2}}}^{2}C_{{12}}S_{{3}}+9/2\,c_{{4}}C_{{12}}+ 6\,{c_{{2}}}^{2}C_{{13}}+45/2 \,c_{{3}}S_{{3}}c_{{2}}+5\,c_{{3}}C_{{13}}+6\,{c_{{2}}}^{3}C_{{12}}+1/2\,c_{{2}}C_{{23}} \big ) \kappa_{11}\,{\kappa_{{2}}}^{3} \nonumber \\
 & & +{b_{{1}}}^{4}  \big ( 9/2\,{c_{{2}}}^{2}{C_{{12}}}^{2}+ 6\,{c_{{2}}}^{2}C_{{13}}+3/2\,c_{{3}}C_{{22}}+39/2 \,c_{{2}}c_{{3}}C_{{12}}+3/2\,c_{{4}}C_{{12}}+9/2\,c_{{3}}S_{{3}}c_{{2}}+3\,{c_{{2}}}^{4}+18\,{c_{{2}}}^{3}C_{{12}} \nonumber \\ 
 & & +  3/2\,c_{{3}}{C_{{12}}}^{2} +15/2\,c_{{2}}c_{{4}}+18\,{c_{{2}}}^{2}c_{{3}}+6\,{c_{{2}}}^{3}S_{{3}} \big ) {\kappa_{11}}^{2}{\kappa_{{2}}}^{2} +{b_{{1}}}^{4} \big ( 9/2\,{c_{{3}}}^{2}+1/2\,c_{{5}}+3/2\,c_{{4}}C_{{12}}+  9\,c_{{2}}c_{{3}}C_{{12}} \nonumber \\ 
 & & +6\,{c_{{2}}}^{2}c_{{3}} \big ) {\kappa_{11}}^{3}\kappa_{{2}}+3/2\,{b_{{1}}}^{4}c_{{2}}c_{{4}}{\kappa_{11}}^{4}  \nonumber \\ 
                           \nonumber
\end{eqnarray}

\noindent where $ c_i\equiv b_i/b_1$ for $i \ge 2$.  In   appendix C,  we list  the expressions for the galaxy correlators up to order $5$. For 
${\bf x}_1={\bf x}_2$, one recovers the expressions of \citet{FG}  (cfr. their eq. 9).\footnote{
Actually we found two typos in their eq. (9). The term  
$210c_{2}^{3}S_{3}$ is actually  $210c_{3}^{2}S_{3}$ and 
$180c_{2}^{2}S_{4}$ should read  $180c_{2}^{3}S_{4}$. Anyway such terms were subsequently neglected in that analysis, leaving
the  author's conclusions unaltered.}  The leading term in eq. (\ref{hiera}) reduces,  in the  one-point limiting case,  to 
\begin{equation}
\kappa_{2,g}=b_1^2\kappa_2.
\label{sigmabias}
\end{equation}

\noindent  We find that, to leading order 
in the product $\kappa_{11} \kappa_2$  the remaining results, $\kappa_{nm,g}$  for $n+m \ge 3$, preserve the hierarchical properties
of matter correlators, i.e. $\kappa_{nm,g}=C_{nm,g} \kappa_{11,g}^{}\kappa_{2,g}^{n+m-2}$, with amplitudes $C_{nm,g}$ given by 

\begin {eqnarray}
C_{12,g}  & = & b_{1}^{-1}\big (C_{{12}}+2\,c_{{2}}\big ) \nonumber  \\
C_{13,g}  & = & b_{1}^{-2}\big (C_{{13}}+3\,c_{{2}}(S_{{3}}+2C_{{12}})+3\,c_{{3}}+6\, c_{{2}}^{2} \big ) \nonumber  \\
C_{22,g}  & = & b_{1}^{-2}\big (C_{{22}}+4\,c_{{2}}C_{{12}}+4\,{c_{{2}}}^{2}\big ) \label{galco} \\
C_{14,g}  & = & b_{1}^{-3}\big (C_{{14}}+4\,c_{{2}}(S_{{4}}+3\,C_{{13}})+12\,(c_{{3}}+3\,c_{{2}}^{2})(S_{{3}}+C_{{12}})+12\,c_{{2}}C_{{12}}S_{{3}}  + 4\,c_{{4}}+36\,c_{{3}}c_{{2}}+24\,c_{{2}}^{3}\big ) \nonumber \\
C_{23,g}  & = & b_{1}^{-3}\big (C_{{23}}+2\,c_{{2}}(C_{{13}}+3\,C_{{22}})+ 3\,(c_{{3}}+c_{{2}}S_{{3}})C_{{12}}+6\,c_{{2}}^{2}(S_{{3}}+3\,C_{{12}})+6\,c_{{3}}c_{{2}} +12\, c_{{2}}^{3} \big ).  \nonumber \\
\nonumber 
\end {eqnarray}
These relations show that in order to draw any conclusions from the galaxy distribution about matter correlations of order $N$, properties of biasing must be specified completely to order $N - 1$.  Note,  also,   that the equations (\ref{galco}) have been  obtained in the large separation approximation and fail as soon as 
$| {\bf x}_1 - {\bf x}_2| <  R$. As a consequence, in the one-point limit they do not converge to  the results of \citet{FG} on the amplitude of the
reduce cumulants, that is 

\begin{eqnarray} 
S_{{3,g}}& =  & b_{1}^{-1}\left(S_{{3}}+3\,c_{{2}}\right)  \nonumber \\ 
S_{{4,g}}& =  &  b_{1}^{-2}\left(S_{{4}}+12\,c_{{2}}S_{{3}}+4\,c_{{3}}+12\,{c_{{2}}}^{2}\right) \label{galmo}  \\
S_{{5,g}}& =  & b_{1}^{-3}\left(S_{{5}}+20\,c_{{2}}S_{{4}}+15\,c_{{2}}{S_{{3}}}^{2}+(30\,c_{{3}}+ 120\,{c_{{2}}}^{2})S_{{3}}+5\,c_{{4}}+60\,c_{{2}}c_{{3}}+60\,{c_{{2}}}^{3}\right). \nonumber \\
\nonumber
\end{eqnarray}

Interestingly,  the hierarchical scaling  (cfr eq. (\ref{hs2})) is not the only matter  property which survives to the local non-linear biasing transformation of eq. \ref{biasfunction}. We also find that  

\begin{eqnarray}
C_{22,g} & = & C_{12,g}C_{12,g}  \nonumber \\
C_{23,g} & = & C_{12,g}C_{13,g}  \\
\label{faco}
\nonumber 
\end{eqnarray}
\noindent that is the reduced galaxy correlators $C_{nm,g}$, conserve the factorization property  of the matter  density field 
(cfr. eq. (\ref{facto}).)

\section{Non linear bias in real space}

The most common methods for estimating the amplitude of the  non-linear bias coefficients $b_i$  rely upon  fitting a theoretical model to  higher order statistical observables, such as the 3-point correlation function \citep[e.g.][]{gaztascoc}, the bispectrum \citep[e.g.][]{Verde}, or the galaxy probability distribution function \citep[e.g.][]{mar05}.

In this section we take a different approach. We explicitly derive analytical  relations directly expressing the non-linear bias coefficients $b_i$ as a function of high order observables.
To this end we cannot simply invert the set of equations \ref{galco} (or \ref{galmo}),   since the $N$th reduced galaxy correlator (or moment) are a function of $N+1$ bias coefficients. We add,  instead,  the expression of the 3rd order  reduced correlator $C_{12,g}$ to the system of equations (\ref{galmo})  and solve the resulting set  
for the biasing coefficients. We obtain 

\begin {eqnarray}
 b_{1} & = & \frac{{\displaystyle 3\,C_{{12}}-2\,S_{{3}}}}{{\displaystyle 3\,C_{{12,g}}-2\,S_{{3,g}}}} \label{biaslin} \\
 b_{2} & = & \frac{{\displaystyle   ( C_{{12}}S_{{3,g}}-S_{{3}}C_{{12,g}}  )  ( 3\,C_{{12}}-2\,S_{{3}}  ) }}{{\displaystyle   ( 3\,C_{{12,g}}-2\,S_{{3,g}}  ) ^{2}}}  \label{biasquad}\\
 b_{3} & =&   ( 9\,S_{{4,g}}C_{{12}}^{2}-12\,S_{{4,g}}C_{{12}}S_{{3}}+4\,S_{{4,g}}S_{{3}}^{2} -12\,S_{{3}}S_{{3,g}}C_{{12}}C_{{12,g}}+24\,S_{{3}}S_{{3,g}}^{2}C_{{12}}  -24\,S_{{3}}^{2}S_{{3,g}}C_{{12,g}} +24\,S_{{3}}^{2}C_{{12,g}}^{2}  \\ 
  & & -9\,S_{{4}}C_{{12,g}}^{2}+12\,S_{{4}}C_{{12,g}}S_{{3,g}}-4\,S_{{4}}S_{{3,g}}^{2}-12\,S_{{3,g}}^{2}C_{{12}}^{2} )   ( 3\,C_{{12}}-2\,S_{{3}}  ) \big / 4 ( 3\,C_{{12,g}}-2\,S_{{3,g}} )^{3} \nonumber \\
 b_{4} & =&  ( -60\,{S_{{3}}}^{3}S_{{4,g}}C_{{12,g}}-108\,S_{{5,g}}{C_{{12}}}^{2}S_{{3}}-120\,S_{{4}}{S_{{3,g}}}^{3}S_{{3}} 
                   -600\,{S_{{3,g}}}^{3}S_{{3}}C_{{12}}^{2}  +495\,S_{{4}}S_{{3}}C_{{12,g}}^{3}-600\,{S_{{3,g}}}^{2}S_{{3}}^{3} C_{12,g}\\
            & & - 270\,S_{{3,g}}S_{{4,g}}C_{{12}}^{3}+72\,S_{{5,g}}C_{{12}}S_{{3}}^{2} +108\,S_{{5}}C_{{12,g}}^{2}S_{{3,g}}
                  +1200\,S_{{3,g}}S_{{3}}^{3}C_{{12,g}}^{2}+600\,S_{{3,g}}^{3}S_{{3}}^{2}C_{{12}}-40\,S_{{4}}S_{{3,g}}^{3}C_{{12}} \nonumber \\ 
            & & +120\,S_{{3,g}}S_{{4,g}}S_{{3}}^{3}-72\,S_{{5}}C_{{12,g}}S_{{3,g}}^{2}-90\,S_{{4}}S_{{3,g}}C_{{12}}C_{{12,g}}^{2}
                  +120\,S_{{4}}S_{{3,g}}^{2}C_{{12}}C_{{12,g}} -930\,S_{{4}}S_{{3,g}}S_{{3}}C_{{12,g}}^{2} \nonumber \\
             & & +580\,S_{{4}}S_{{3,g}}^{2}S_{{3}}C_{{12,g}} 
            +630\,S_{{3,g}}S_{{4,g}}C_{{12}}^{2}S_{{3}} -480\,S_{{3,g}}S_{{4,g}}C_{{12}}{S_{{3}}}^{2}+180\,S_{{3,g}}^{2}S_{{3}}C_{{12}}^{2}C_{{12,g}}
                  -600\,S_{{3,g}}^{2}S_{{3}}^{2}C_{{12}}C_{{12,g}} \nonumber \\ 
                   & & +270\,S_{{3,g}}S_{{3}}^{2}C_{{12}}C_{{12,g}}^{2}
                     -135\,S_{{3}}S_{{4,g}}C_{{12}}^{2}C_{{12,g}}+180\,S_{{3}}^{2}S_{{4,g}}C_{{12}}C_{{12,g}} +54\,S_{{5,g}}C_{{12}}^{3}-16\,S_{{5,g}}S_{{3}}^{3} -54\,S_{{5}}C_{{12,g}}^{3} \nonumber \\ 
                      & & +16\,S_{{5}}S_{{3,g}}^{3} + 240\,S_{{3,g}}^{3}C_{{12}}^{3}-690\,S_{{3}}^{3}C_{{12,g}}^{3} ) ( 3\,C_{{12}}-2\,S_{{3}}  ) \big / 10 ( 3\,C_{12,g}-2\,S_{3,g} )^{4}. \label{biasfact} \nonumber \\
                        \nonumber
\end{eqnarray}

This set of equations allow us to investigate the eventual non-linear  character of the biasing function up to  order $4$ by exploiting information encoded in the reduced correlators up to  order $3$  and the reduced moments up to  order $ \leq 5$.   In a forthcoming  analysis we will investigate up to what
precision the coefficients $b_i$ can  be estimated using data of large redshift surveys such as BOSS, BigBOSS and  EUCLID. 
Note that,  if we set $\beta_R(r)=0$ in eq. (\ref{c12prediction}),  then our expressions  for $b_1$ and $b_2$  (cfr. should eqs. (\ref{biaslin}) and (\ref{biasquad})) 
reduce to equations  (4) and (5) originally derived by  \citet{Szapudi98}. As stressed by this author, in this formalism biasing coefficients are not anymore simple  parameters to be estimated  (by maximizing, for example,  the likelihood of observables that are sensitive to them  such as the  reduced skewness $S_3$ \citep{gazta94, gaztafrie94},  the bispectrum \citep{Fry94, scoc98, feld01, Verde}, the 3-point correlation function \citep{gaztascoc, panszapudi} or the full probability distribution function of the density fluctuations \cite{mar05,mar08},   but they become themselves estimators.

In what follows we will focus our analysis on the linear biasing parameter $b_1$ only. In our formalism, this quantity is  explicitly expressed   
in terms of  the amplitude of third-order statistics (i.e. the reduced cumulants and correlators) and it is independent from the amplitude of second-order statistics such as 
$\sigma_R$ and $\xi_R$. Notwithstanding, $b_1$ seems to depend  on the shape of 
the power spectrum of the matter density fluctuations via the terms  $\gamma_R$ and  $\beta_R(r) $ that appears 
in eq.  (\ref{c12prediction}).  We will demonstrate in the next session  that  the  inclusion of the correction term  $\beta_R(r)$ 
in our analysis  has the additional advantage of  making  the linear bias coefficient $b_1$ effectively independent from any assumption 
about second order statistics, i.e.  independent not only  from the amplitude but also from the shape  of the linear matter power spectrum.  
This {\it fully third-order} dependence of the linear biasing parameter estimator $b_1$ allows us to construct a consistent {\it second-order}  estimator of the matter density fluctuations  $\sigma_R$  on a given linear scale $R$. 

\section{The $rms$ fluctuations  of the linear mass density field}

Now that all the ingredients are collected, we detail how we construct our estimators of the linear matter density fluctuations $\sigma_R$.
From eq. (\ref{s3prediction}), (\ref{c12prediction}),  (\ref{sigmabias}) and  (\ref{biaslin}) we obtain 

\begin{equation}
\sigma_R= \frac{\tau_{g,R}(r)}{\beta_R(r) - \gamma_R}  \sigma_{g,R} 
\label{s1}
\end{equation}
\noindent where 

\begin{equation}
\tau_{g,R}(r)=3C_{12,g,R}(r)-2S_{3,g,R}
\end{equation}
and where the suffix $g$ indicates that the relevant quantities are evaluated using data.  For clarity, the scale dependence of  third order 
statistics is explicitly highlighted.  Apparently,  the right-hand side of the above equations depends on the overall shape of the {\it a-priori}
unknown matter power spectrum. In reality, the terms $\beta_R(r)-\gamma_R$ can be consistently estimated from observations without  any additional theoretical assumption. 
To show this, we define

\begin{equation}
\alpha_R(r) \equiv \frac{d\log \eta_R(r)}{d \log R}\end{equation}
\noindent where 

\begin{equation}
\eta_R(r) \equiv \frac{\xi_R(r)}{\sigma^2_R}.
\end{equation}

\begin{figure}
   \centering
   \includegraphics[width=8.5cm]{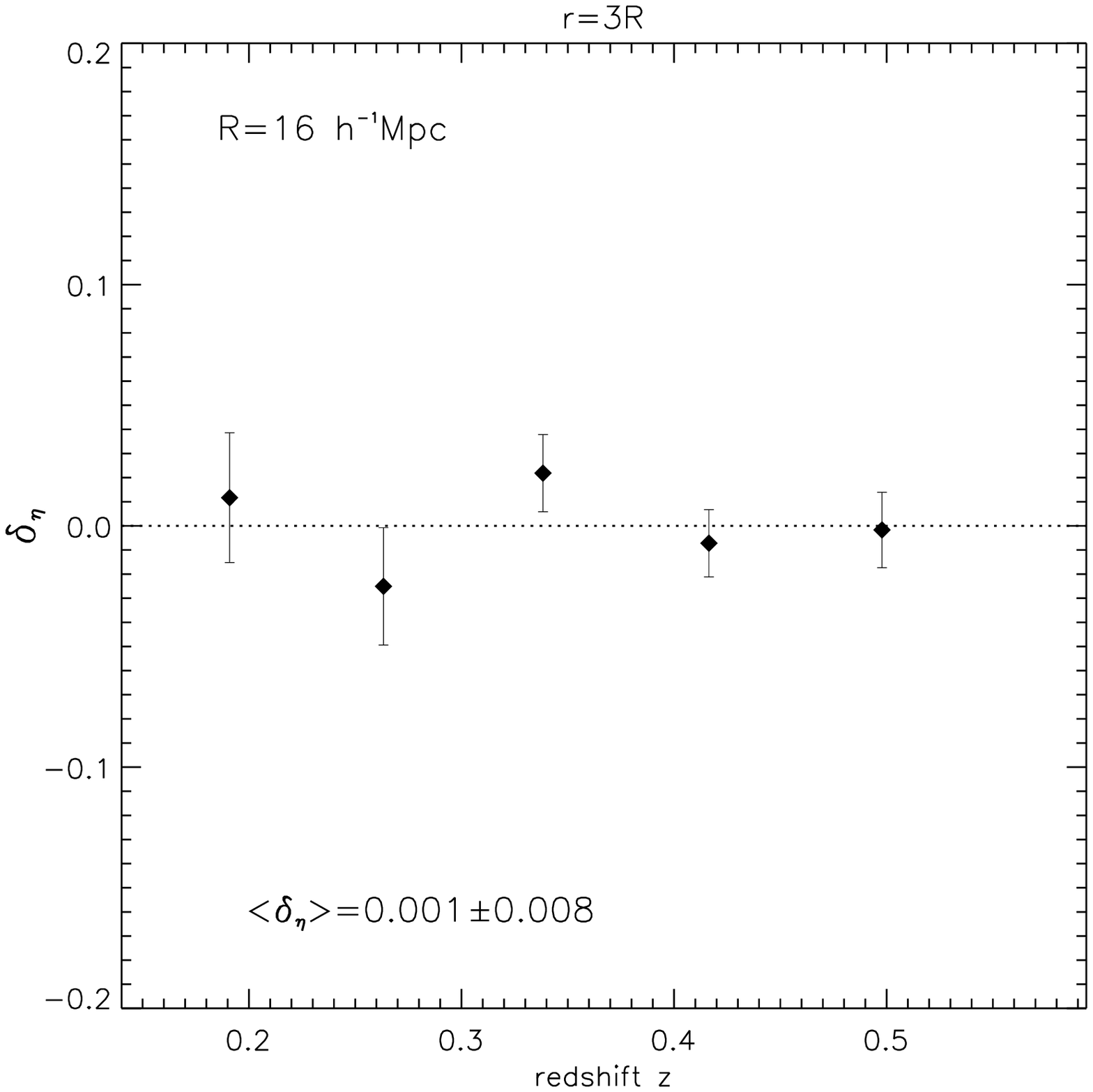}
   \includegraphics[width=8.5cm]{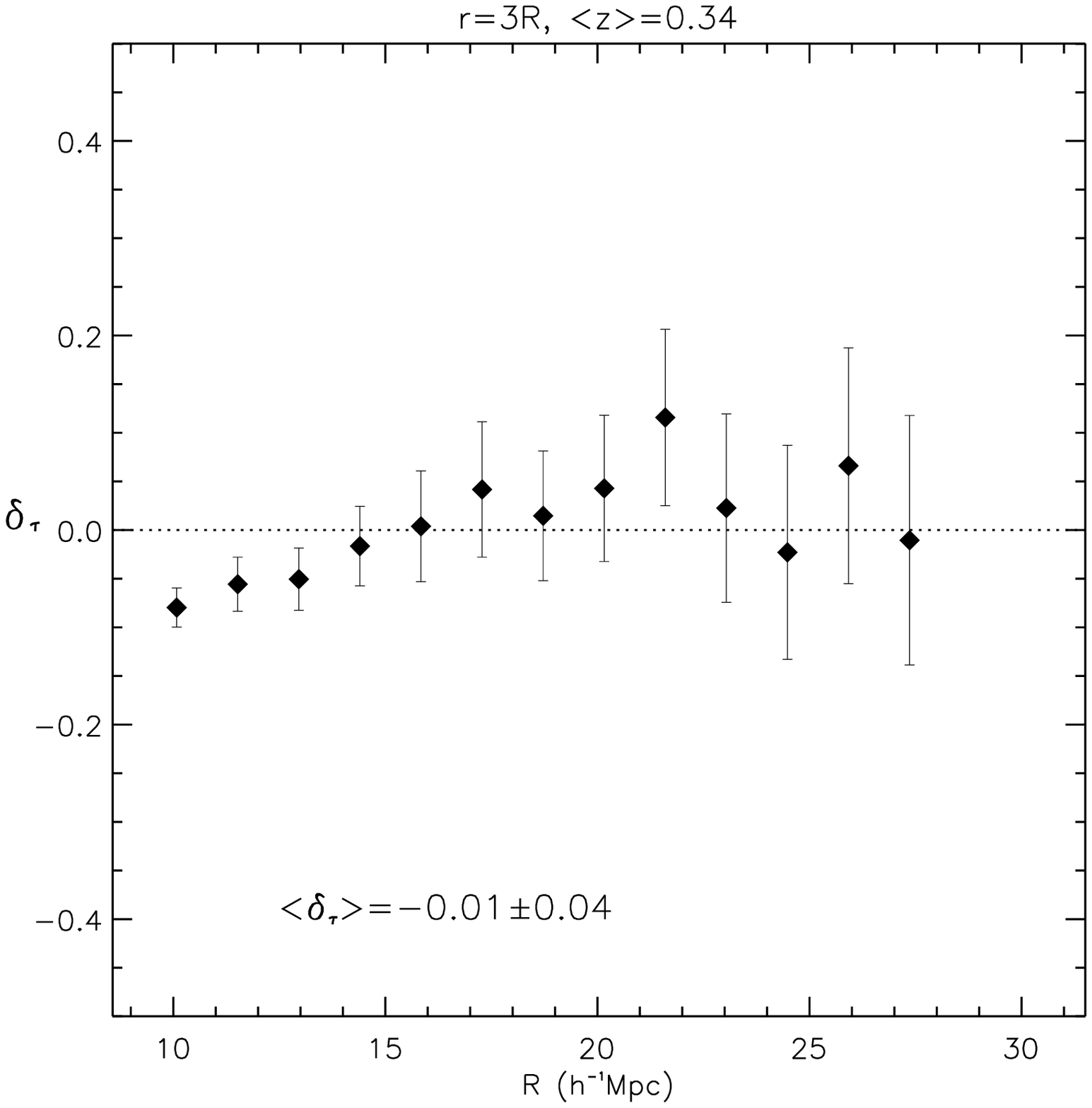}
   \caption{{\it Left:} the relative error  arising from estimating the $\eta$-function using  galaxies instead of matter. The inaccuracy ${\displaystyle \delta_{\eta}=\eta_{g,R}^z / \eta_{R}^{\;} -1}$ is evaluated  at different epochs and is determined by assuming $R= 16 h^{-1}$Mpc and $r=3R$.   
    Simulated  data are extracted from the mock  catalogs  described in section \S 6.2.
   {\it Right:} relative difference between the real-  and redshift-space estimation  of  $\tau_{g,R}\equiv 3C_{12,g}(r)-2S_{3,g}$.  The inaccuracy $\delta_{\tau}=
   \tau_{g,R}^z(r) / \tau_{g,R}^{\;}(r) -1$ is plotted as a function of the smoothing scale $R$ and is computed at a  separation $r=3R$.
   The dotted line represents the case in which  computing the $\tau$ statistics   using observed redshifts is equivalent to using cosmological redshifts. 
   Real- and redshift-space simulated data are extracted from the mock  catalogs  described in section \S  6.2.
  In  both panels the error bars represent $1 \sigma$ standard deviation.}               
   \label{taugrz}
\end{figure}

\noindent As far as matter particles are considered, the previous definitions imply that  $\alpha_R(r)=\beta_R(r)-\gamma_R$.  By combining the expression
for  $\xi_{g,R}$ given by eq.   \ref{hiera} with  its one-point limiting case i.e. 

\begin{equation}
\sigma_{g,R}^2(z)=b_1^2\sigma_R^2(z)\left\lbrace 1+\left(1/2{c_2}^2+S_{3,R} c_2+c_3\right) \sigma_R^2(z)\right\rbrace,
\label{kg2}
\end{equation}

\noindent and  using the fact that, on scales where WNLPT applies,  $\sigma_R^2 \ll 1$,   we obtain 

\begin{equation}
\eta_{g,R}(r) \sim \eta_R(r)-\left\lbrace(S_{3,R}-C_{12,R})c_2+1/2{c_2}^2\right\rbrace\xi_R(r)+1/2{c_2}^2\eta_R(r)\xi_R(r),
\label{etag}
\end{equation}

\noindent where  $\eta_{g,R}(r)= \xi_{g,R}(r)/ \sigma^2_{g,R}$ and where  the terms on the RHS have been sorted by order of magnitude.
Finally, in the  LS limit [$\xi_R(r)$ is negligible with respect to  $\eta_R(r)$],  
the above equation reduces to $\eta_{g,R}(r) \sim \eta_R(r)$. The level of accuracy of this  approximation 
is presented  in Figure  \ref{taugrz} where we show  that,  on a typical  scale  ($R\sim 16h^{-1} $ Mpc), the 
imprecision  is less than $0.5\%$  at any cosmic epoch  investigated  ($0<z<0.6$).  Since
$\alpha_{g, R}(r) = \alpha_R(r)$, we obtain 

\begin{equation}
\sigma_R= \frac{\tau_{g,R}(r)}{\alpha_{g,R}(r)} \sigma_{g,R}
\label{srealspace}
\end{equation}

\noindent where $\alpha_{g,R}(r)=d\log \eta_{g,R}(r)/d \log R$. 

Counts-in-cells  techniques provide and estimate of the terms on the RHS  of eq. \ref{srealspace}. 
In this regard,  a central point worth stressing concerns the continuum-discrete connection.  Biasing is not the only obstacle hampering the 
retrieval of matter properties from the analysis of galaxy catalogs. The formalism needs also to correct for the fact that  the
galaxy distribution  is  an intrinsically discrete process. These issues will be thoroughly addressed in Section \S 6.1 where we present the strategy that  we have adopted in order to minimize
the sampling noise.

Redshift space distortions are an additional effect   that needs modeling.
Results presented in the previous sections  strictly hold in real (configuration) space.  Therefore, 
the feasibility of extracting the value of mass fluctuations $\sigma_R $  via  eq. (\ref{srealspace})  rests upon the possibility of expressing 
real space variables ($b_1^{-1}=\tau_{g,R}/ \alpha_{g,R}$,    and   $\sigma_{g,R}$) in terms of redshift space observables.
On the  large (linear) cosmic scales where our formalism applies, the Kaiser model \citep[e.g.][]{Kaiser}  effectively describes  the mapping between real and redshift space expressions of second order statistics. The transformation is given by 

\begin{equation}
\sigma_{g,R}^z=\left\lbrack1+\frac{2}{3}\frac{f}{b_1}+\frac{1}{5}\left(\frac{f}{b_1}\right)^2\right\rbrack^{1/2} \sigma_{g,R},
\label{kaisercorr}
\end{equation}
where the suffix $z$   labels measurements in  redshift  (as opposed to configuration)  space, and where 
$f$ is the logarithmic derivative of the linear growth factor $D(a)$ with respect to the scale factor $a$. 
Much easier is the transformation rule for $\alpha_{g,R}$: in linear regime it is unaffected by redshift distortions, 
that is  $\alpha^z_{g,R}=\alpha_{g,R}$.

Assessing the  impact of peculiar motions  on third order statistics is less straightforward. Up to now all the formulas were 
derived analytically from theory.  To address this last issue we now use numerical simulations. 
By running  some tests using a suite of simulated galaxy catalogs (described in section \S 6.2) we conclude  that  the amplitude of 
$S_{3,g}$ and $C_{12,g}$  are  systematically (and non-negligibly) higher in $z$-space and that the relative  overestimation 
systematically increases as a function of the order of the statistics considered (see  Figures \ref{d3bigboss} and  \ref{highcor} in section \S 6.2). 
Notwithstanding, from the theoretical side, the expressions of third order statistics 
($S_{3,R}=S_3 + \gamma_R$ and $C_{12}=(2S_3)/3 +\gamma_R/3+\beta_{R}(r)/3$) imply
that the linear combination  $3C_{12,R}-2S_{3,R}$ should be much more insensitive to redshift distortions. 
Note, in particular,  that both $\gamma_R$ and $\beta_R(r)$ are unaffected by linear motions.
However convincing it might seem, this guess applies only to matter particles.  In order to draw definitive  conclusions about the impact 
of peculiar motions on  $\tau_{g,R}$, we have used   N-body galaxy simulations.  Guided  by synthetic catalogs we  demonstrate  
that the biased galaxy statistics $\tau_{g,R}$ is effectively unaffected by  redshift distortions. This conclusion  is graphically presented in 
Figure \ref{taugrz} where we  show that the relative error introduced  by reconstructing the statistics using observed redshifts, 
instead of the cosmological ones, is progressively smaller as $R$ increases, and it is   globally  $<\sim 2\%$. 
  
By incorporating these results into the formalism we finally  obtain 

\begin{equation}
{\displaystyle \hat{\sigma}_{R}}=\sigma^z_{g,R}  \Bigg [ \bigg(\frac{\alpha^{z}_{g,R}}{\tau^{z}_{g,R}} \bigg)^2+\frac{2}{3} \frac{\alpha^z_{g,R}}{\tau^z_{g,R}} f +\frac{1}{5} f^2 \Bigg ] ^{-1/2} 
\label{sigestimz}
\end{equation}

\noindent an estimator  that is manifestly independent from any assumption about the amplitude and shape of the linear matter power spectrum.
Also,  this formula is independent from  any assumption about the value of  the Hubble constant  $H_0$.  
Only an {\it a-priori} gravitational  model must be assumed to correct for redshift  space distortions in the local universe, that is to evaluate the growth rate function  $f(z)$.
This  introduces an additional strong dependence on the  cosmological parameters $\Omega_M$ and $\Omega_{\Lambda}$  on top 
of the marginal one that is  forced upon  when we compute metric distances in order to estimate $\sigma^z_{g,R}$, $\alpha^{z}_{g,R}(r)$ and $\tau^{z}_{g,R}$.
More quantitatively, when the input cosmological parameters are chosen in the parameter plane delimited by $0 \leq \Omega_M \leq 1$ and $0 \leq \Omega_{\Lambda} \leq 1$ , the maximum relative variation of the estimates  with respect to their fiducial value in the $\Lambda$CDM cosmological model  are max$|df/f|\sim 0.6$, max$|d\tau/\tau|\sim 0.2$, max$|d\alpha/\alpha|=0.3$, 
and max$|d\sigma/\sigma|=0.15$. The weak cosmological dependence of $\tau^{z}_{g,R}$ follows from the fact that neither the  reduced cumulants  nor the reduced  correlators  of mass are effectively sensitive to the background cosmology.   We expect them to be essentially unaffected also by  sensible modifications of 
the gravitational theory (Gazta\~naga \& Lobo (2001), Multamaki et al. (2003),  but see for example Freese \& Lewis (2002) or  Lue,  Scoccimarro \& Starkman  (2001)
for more radical scenarios where this expectation is not met).
We have also noted that the denser in matter is the cosmological model the more overestimated are the values of both of $\alpha^z_{g,R}$ and $\tau^z_{g,R}$. Since also  the amplitude of the cosmological dependence  is nearly the same,  we expect the ratio $\alpha^z_{g,R}/\tau^{z}_{g,R}$, i.e. the linear bias parameter, to be nearly insensitive  to the underlying  cosmological model. We have quantitatively verified this statement  in Figure 11 of section \S 6.5. In conclusion, the estimator in  eq. (\ref{sigestimz}) is 
sensitive to cosmology mainly through the  growth rate function $f(z)$ and, to some degree,  through $\sigma^{z}_{g,R}$.

Once the cosmological background is  known  via  independent techniques (e.g \cite{ast06, koma,  mar10})) the strategy we have outlined 
offers the possibility to estimate in  a direct way  the amplitude and time evolution of matter fluctuations.  
The formalism could also be implemented,  in a reverse direction,  to  probe  the coherence of the gravitational  instability paradigm. 
Any eventual discrepancy resulting from the comparison  of the measurements  (cfr.  eq.   (\ref{sigestimz})) with theoretical predictions 
 (cfr.  eq.  (\ref{teosigma})) provides evidence that either the assumed set of cosmological parameters are wrong, either the assumed  power spectrum of matter 
fluctuations is poorly described by linear theory, either the time dependence of the linear growing mode $D(t)$  is deduced in the context of an improper gravitational model. 
This last testing modality will be explored  in a further paper.

\section{Applying the Method}
The practical implementation of the method, including successful tests of its robustness, is  discussed in this section.   
We first present a strategy to  estimates the correlators of order $(n,m)$ of  discrete 3D  density fields such as those sampled
by  galaxy redshift surveys.  We then show that, by applying the test formalism  to N-body simulations of the large scale structure of the universe, 
we are able to recover the amplitude and scaling of the linear matter fluctuations $\sigma_R(z)$. A strategy to test the coherence of the results 
and to validate our conclusions  is also designed, applied and  discussed.

\subsection{Statistical estimators of the galaxy correlators} 

The galaxy distribution is  a discrete, 3-dimensional  stochastic process. The random variable $N$  models the number of galaxies  within  typical
cells (of constant comoving size)  that ideally tesselate  the universe or,  less emphatically,  a given redshift survey. Notwithstanding, a variable that is more directly linked to theoretical predictions of cosmological perturbation theory is the adimensional galaxy excess 
\begin{equation}
\delta_N\equiv\frac{N}{\bar N}-1,
\end{equation}
where $\bar N$ is the mean number of galaxies contained in the cells.

To estimate one-point moments  of the galaxy overdensity field ($\mu_{n,g}= \langle\delta_N^n\rangle$),  we fill the survey volume with the maximum number ($N_t$) of non-overlapping spheres of radius $R$ (whose center is called {\it seed}) and we compute 
\begin{equation}
\hat\mu_{n,g}=\frac{1}{N_t}\sum_{i=1}^{N_t}\delta_{N,i}^n,
\end{equation}

\noindent where $\delta_{N,i}$ is the adimensional counts excess  in the $i$-th sphere. 

As far as two-point statistics are concerned, we add a motif around each previously positioned spheres. The center of each new sphere is separated from the seed 
by the length $r=nR$ and the pattern is designed in such a way to maximize the number of quasi non-overlapping spheres at the given distance $r$ 
(the maximum  allowed  overlapping  between contiguous spheres  is $2\%$ in volume.)
The moments  $\mu_{nm,g}=  \langle\delta^n_{N,i}\delta^m_{N,j}\rangle$, that is the average of the  excess counts over all the $i$ and $j$  spherical cells 
at separation $r$,  are  estimated as 
\begin{equation}
\hat\mu_{nm,g}=\frac{1}{2N_tN_{mot}}\left\lbrace\sum_{i=1}^{N_t}\delta_{N,i}^n\sum_{j=1}^{N_{mot}}\delta_{N,j}^m+\sum_{i=1}^{N_t}\delta_{N,i}^m\sum_{j=1}^{N_{mot}}\delta_{N,j}^n\right\rbrace,
\end{equation}

\noindent where  $N_{mot}$ is the number of spheres at distance $r$ from a given seed, and where we have assumed that the stochastic process is stationary.
The expression of the one- and two-point cumulant moments $k_{nm,g}$  follows immediately  from eq. (\ref{expa}). 
 
Since galaxies counts are a discrete sampling of the underlying continuous stochastic field $\lambda_g({\bf x})$ (see section \S 2.1)  
it is necessary to correct our estimators for discreteness effects. In other terms, the quantity of effective physical  interest that we want to estimate 
is   $\delta_{g,R}({\bf  x})\equiv \Lambda_g({\bf x}) / {\bar \Lambda_g}-1$ where $\Lambda_g({\bf x}) =\int_{V({\bf x})} \lambda({\bf x'})d^3x'$ is the 
continuous limit of the discrete counts  N in the volume V. To this purpose, following standard practice in the field, we model the sampling as a  local Poisson process  (LPP,  \citet{Layser}) 
and we  map moments of the discrete variable N into moments of its continuous limit by using 

\begin{equation}
\langle\Lambda_g^n\rangle=\langle N(N-1)...(N-n+1)\rangle=\langle(N)_f^n\rangle,
\label{unpoint}
\end{equation}

\noindent in the case of  one-point statistics,  and its generalization \citep{sza2}

\begin{equation}
\langle\Lambda_g^n({\bf  x}_1)\Lambda_g^m({\bf  x}_2)\rangle=\langle(N_1)_f^n(N_2)_f^m\rangle,
\label{deuxpoint}
\end{equation}

\noindent for the  two-point  case. As a result, the  estimators of  $ k_{nm,g}=\langle\delta^n_{g,R}(\vec x)\delta^m_{g,R}(\vec x+\vec r)\rangle_c$ 
corrected for shot noise effects are \citep{sza1, Angulo}

\begin{eqnarray}
\hat k_{2,g} &= &\hat\mu_{2,g}-\Nb^{-1} \label{ccum}\\
\hat k_{11,g} &=   &\hat\mu_{11,g} \nonumber \\
\hat k_{3,g} &= &\hat\mu_{3,g}-3\Nb^{-1}\hat\mu_{2,g}+2\Nb^{-2} \nonumber \\
\hat k_{12,g} &= &\hat\mu_{12,g}-\Nb^{-1}\hat\mu_{11,g} \nonumber  \\
\hat k_{4,g} &= &\hat\mu_{4,g}-3\hat\mu_{2,g}^2-6\Nb^{-1}\hat\mu_{3,g}+11\Nb^{-2}\hat\mu_{2,g}-6\Nb^{-3}\nonumber  \\
\hat k_{13,g} &= &\hat\mu_{13,g}-3\hat\mu_{2,g}\hat\mu_{11,g} +2\Nb^{-2}\hat\mu_{11,g} -3\Nb^{-1}\hat\mu_{12,g} \nonumber \\
\hat k_{22,g} &= &\hat\mu_{22,g}-2\hat\mu_{11,g}^2-\hat\mu_{2,g}^2+\Nb^{-2}\hat\mu_{11,g}-2\Nb^{-1}\hat\mu_{12,g}\nonumber \\
\hat k_{5,g} &= &\hat\mu_{5,g}-10\hat\mu_{2,g}\hat\mu_{3,g}-10\Nb^{-1}\lbrace\hat\mu_{4,g}-3{\hat\mu_{2,g}}^2\rbrace+ 35\Nb^{-2}\hat\mu_{3,g}-50\Nb^{-3}\hat\mu_{2,g}+24\Nb^{-4}\nonumber \\
\hat k_{14,g} & = &\hat\mu_{14,g}-6\hat\mu_{12,g}\hat\mu_{2,g}-4\hat\mu_{11,g}\hat\mu_{3,g}-6\Nb^{-1}\lbrace\hat\mu_{13,g}-3\hat\mu_{11,g}\hat\mu_{2,g}\rbrace+ 11\Nb^{-2}\hat\mu_{12,g}-6\Nb^{-3}\hat\mu_{11,g}  \nonumber \\
\hat k^g_{23,g} & = &\hat\mu_{23,g}-6\hat\mu_{11,g}\hat\mu_{12,g}-3\hat\mu_{2,g}\hat\mu_{12,g}-\hat\mu_{2,g}\hat\mu_{3,g}-3\Nb^{-1}\lbrace\hat\mu_{22,g}-2{\hat\mu_{11,g}}^2-{\hat\mu_{2,g}}^2\rbrace -\Nb^{-1}\lbrace\hat\mu_{13,g}-3\hat\mu_{11,g}\hat\mu_{2,g}\rbrace+5\Nb^{-2}\hat\mu_{12,g}+ \nonumber \\ 
& & -2\Nb^{-3}\hat\mu_{11,g  }\nonumber 
\end{eqnarray}

\noindent where we have set  $\hat\mu_{0i,g}=\hat\mu_{i0,g}\equiv \hat\mu_{i,g}$ and $\hat k_{0i,g}=\hat k_{i0,g} \equiv \hat k_{i,g}$.

\begin{figure}
\begin{center}$
\begin{array}{cc}
\includegraphics[width=8cm]{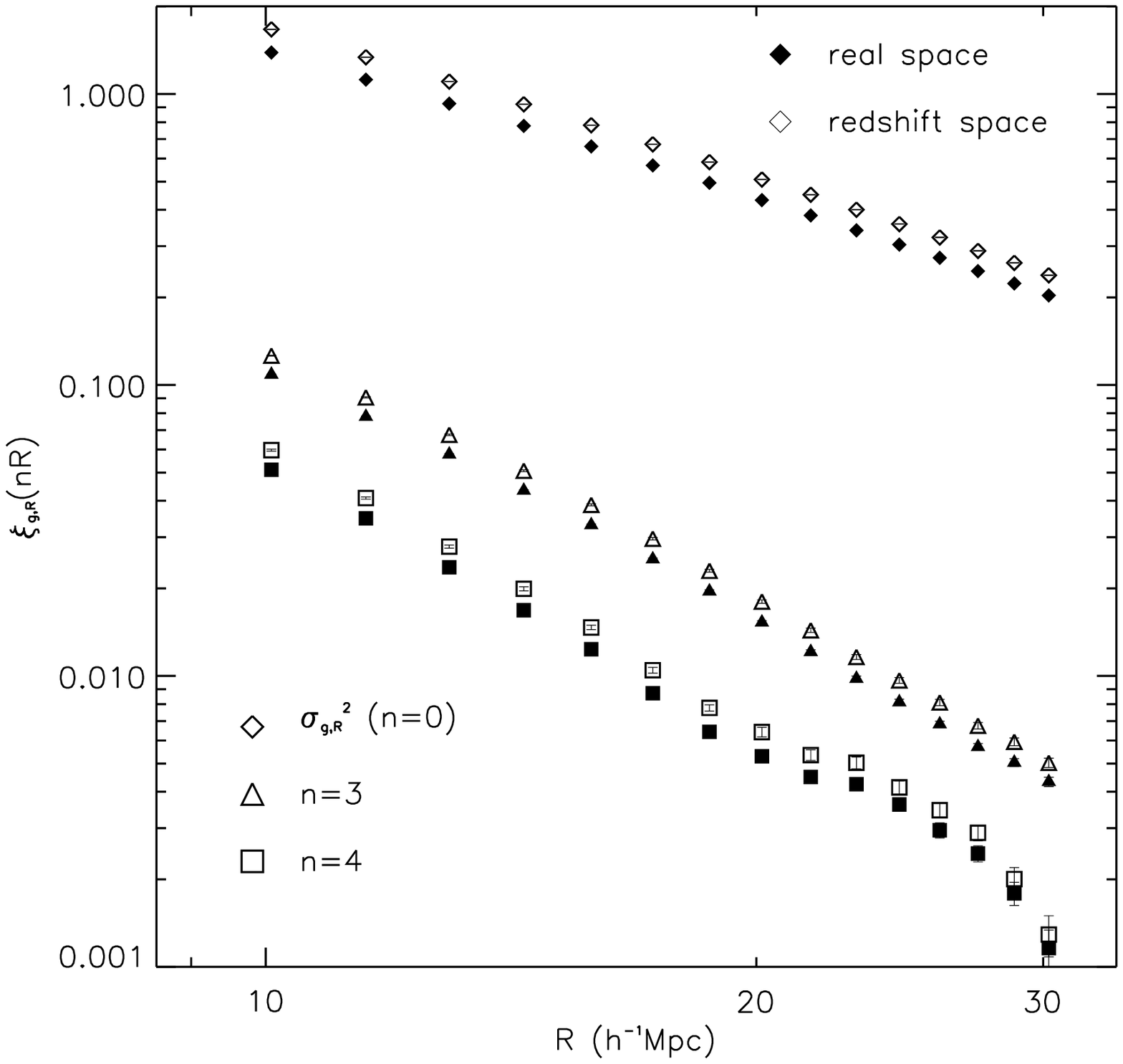} &
\includegraphics[width=8cm]{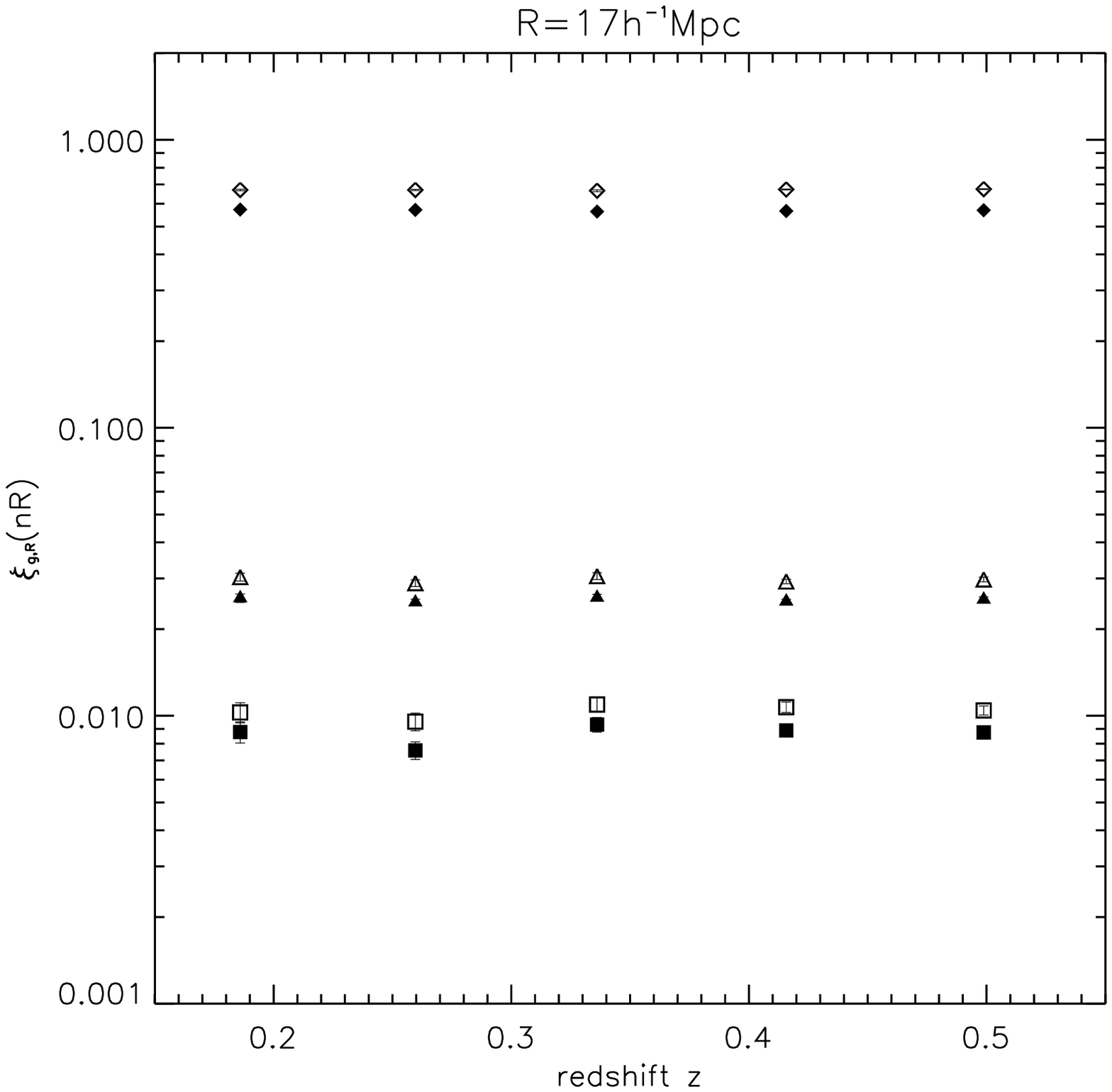}
\end{array}$
\end{center}
\caption{\small {\it Left:} the correlator of order $(1,1)$ of  the smoothed galaxy overdensity field at the average redshift $z=0.34$ is shown as a function of the smoothing radius $R$, and for different values of the correlation length ($n=\{0,3,4\}$). In the degenerate case $n=0$  we recover the variance of the galaxy fluctuation field.  {\it Right: } the redshift evolution of $\xi_{g,R}(nR)$ at a given smoothing scale ($R=17h^{-1}$Mpc)  is shown for   different values of the correlation length ($n=\{0,3,4\}$).  The statistics are computed in both real (solid symbols) and redshift space (unfilled symbols). Each point is the average of the results obtained from  8 independent full-sky LRGs catalogs. Errorbars are estimated  as the standard  error of the mean.}               
   \label{d2bigboss}
\end{figure}

\begin{figure}
\begin{center}$
\begin{array}{cc}
\includegraphics[width=8cm]{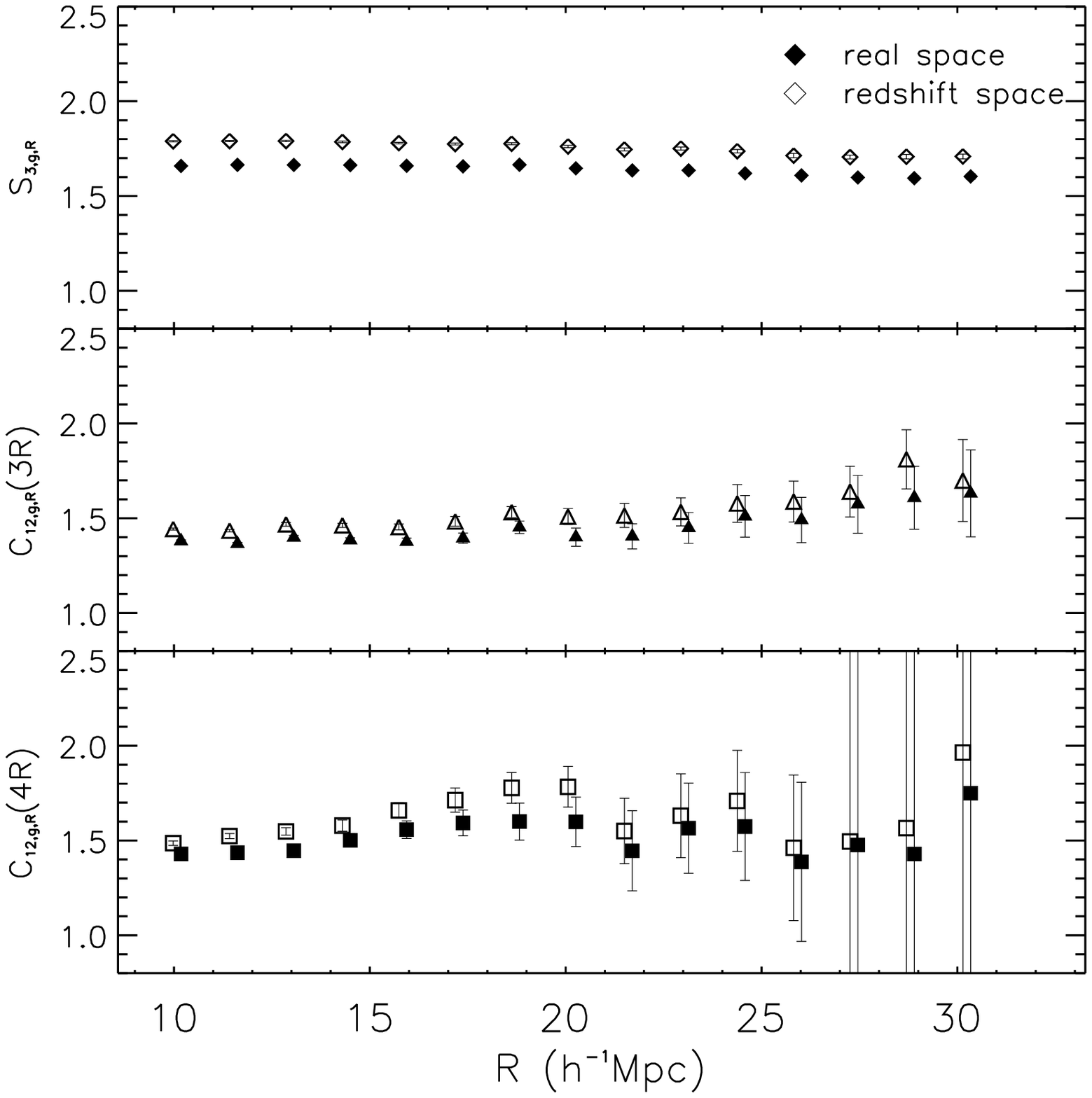} &
\includegraphics[width=8cm]{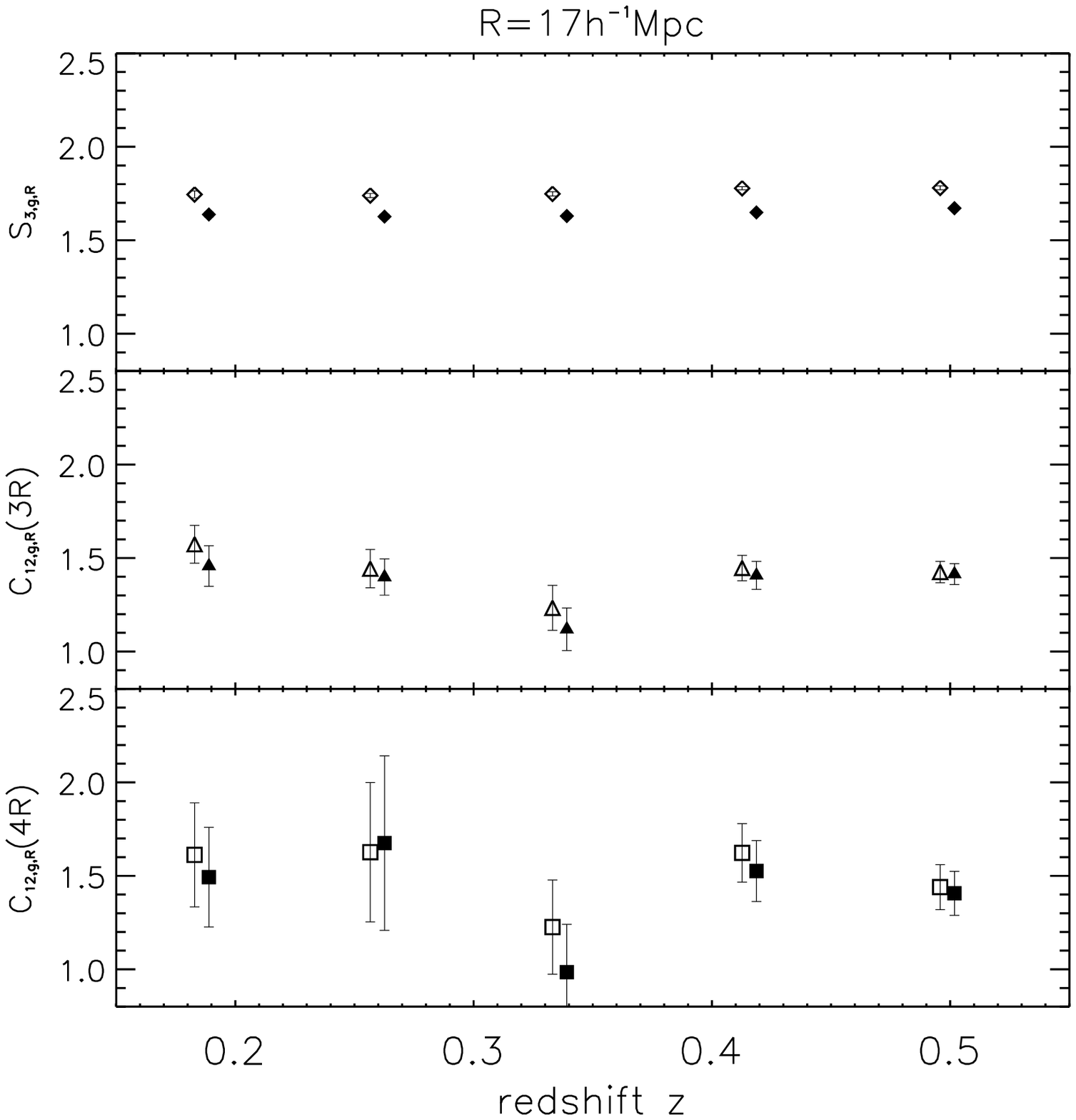}
\end{array}$
\end{center}
\caption{\small {\it Left:} the reduced  correlator of order $(1,2)$ of  the smoothed galaxy overdensity field at the average redshift $z=0.34$ is shown as a function of the smoothing radius $R$, and for different values of the correlation length ($n=0$ (upper panel) $n=3$ (central panel) and $n=4$ (lower panel)). In the degenerate case $n=0$  we recover the reduced skewness of the galaxy fluctuation field.  {\it Right: } the redshift evolution of $C_{12,g,R}(nR)$ at a given smoothing scale ($R=17h^{-1}$Mpc)  is shown for   different values of the correlation length ($n=\{0,3,4\}$).  The statistics are computed in both real (solid symbols) and redshift space (unfilled symbols). Each point is the average of the results obtained from  8 independent full-sky LRGs catalogs. Errorbars are estimated  as the standard  error of the mean.}               
   \label{d3bigboss}
\end{figure}

\begin{figure}
\begin{center}
\includegraphics[width=10cm]{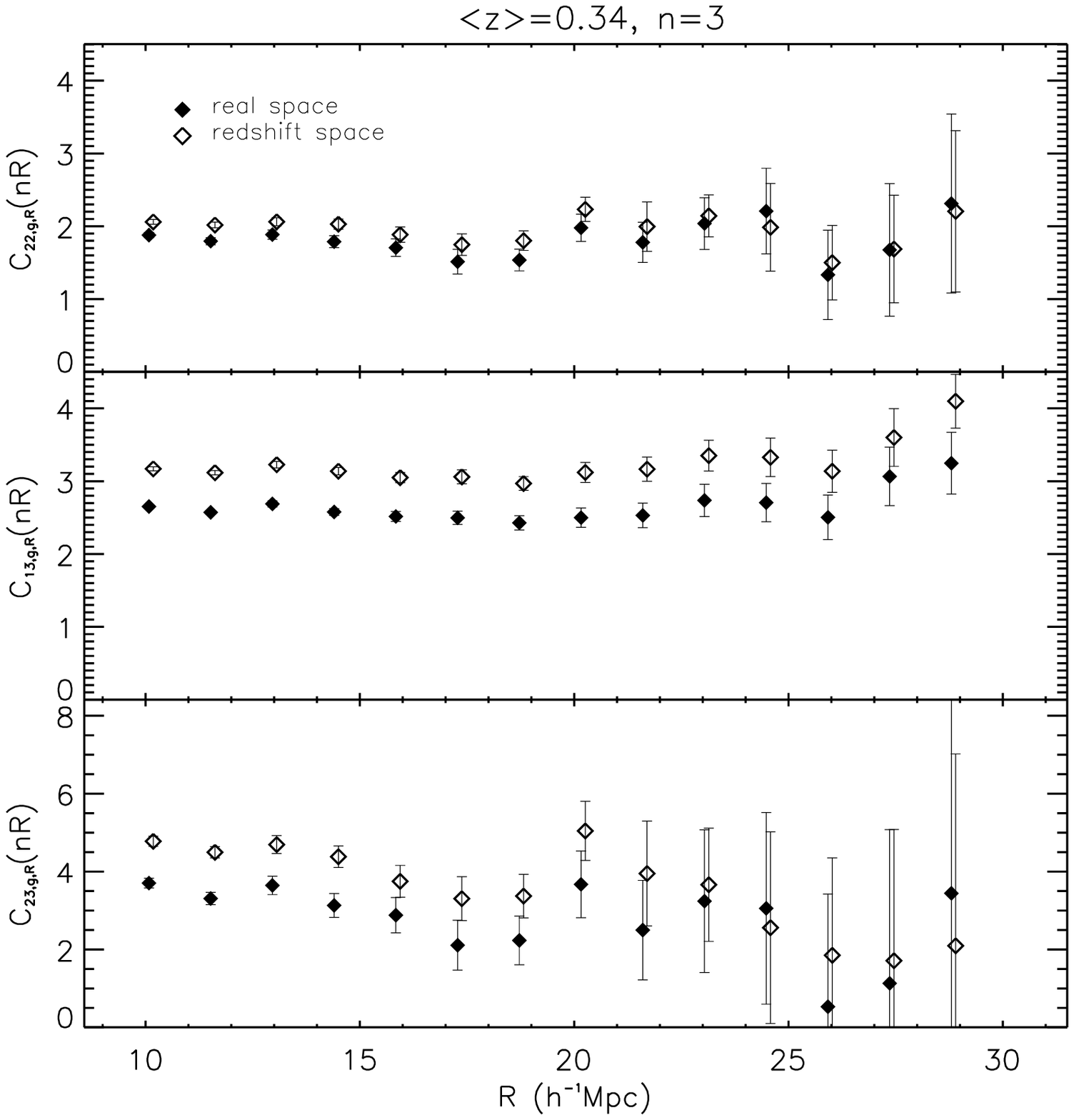}
\end{center}
\caption{\small Tthe reduced galaxy  correlators of order 4 ($C_{13,g,R}$ and $C_{22,g,R}$) and of order 5  ($C_{23,g,R}$) at the average redshift $z=0.34$ and at the correlation separation $r=3R$ are shown as a function of the smoothing radius $R$. The statistics are computed in both real (solid symbols) and redshift space (unfilled symbols). Each point is the average of the results obtained from 8  independent LRGs catalogs. Errorbars are estimated  as the standard  error of the mean.}               
\label{highcor}
\end{figure}

\subsection{High order correlators extracted from cosmological simulations}

We use  numerical experiments   simulating the spatial clustering of  Luminous Red Galaxies (LRGs)  of the Sloan Digital Sky Survey (SDSS) to validate the method,  to test its end-to-end coherence and  to spot the presence of eventual systematics.
We have run the whole pipeline on two different simulations, namely Horizon  \citep{Kimetal, Dubinskietal, KP} and Las Damas \citep{mcb}. Ideally, this way,  we are the least dependent upon 
the specific simulation strategy and technique. Since  the outcome and salient features of the analysis  are  essentially the same,  we here only present the analysis of the Horizon  simulations.
This is a large simulation  ($4120^3$  particles in the box) in which  LRGs galaxies are selected by finding the most massive gravitationally bound, cold dark matter halos. It is characterized by the following set of cosmological assumptions  $(\Omega_m=0.26, \Omega_\Lambda=0.74, w_0=-1, w_a=0, H_0=72$km/s/Mpc,$  \Omega_b=0.044, n_s=0.96, \sigma_8=0.79)$. We have analyzed 8  nearly independent, full-sky light cones extending over the interval  $0.15<z<0.55$, each covering a volume of 
$13\;h^{-3}$Gpc$^3$ and containing nearly $3.8\cdot 10^6$  LRGs galaxies. A mass threshold decreasing with redshift was chosen so  to force Horizon's comoving number density profile to be constant with redshift
and reproduce the density profile observed in the SDSS LRGs sample. As a  consequence,  the simulated sample can be considered with good approximation as being volume limited. The mean comoving number density  is $\bar n=3.0\times 10^{-4}\;h^3$Mpc$^{-3}$ and the mean inter-galaxy separation is $l \sim 19\;h^{-1}$Mpc.  As an example, the average number of LRGs galaxies (${\bar N}$) inside a spherical cell of radius $R=(10,15,20, 25) h^{-1}$Mpc is approximately  $1,5, 10$ and $20$.

In the left panel of Figure \ref{d2bigboss} we show the correlation function of the smoothed galaxy overdensity field $\xi_{g,R}=\langle \delta_{g,R}({\bf x})\delta_{g,R}({\bf x}+{\bf r})\rangle_c$ for different values of the correlation length ($r=nR$) in both real and redshift spaces.  
Note that in the degenerate case $n=0$ we recover  the variance of the  galaxy density fluctuations on a scale $R$. If the cell separation increases,  the  amplitude of the galaxy correlators of order 2 decreases. Their  $R$-scaling  is nearly similar to the slope of the analogous  statistics computed for the  matter density  field (see Figure \ref{betatheo}) with a slope of $\sim -2.25$ at $R=10h^{-1}$Mpc and $\sim -3.4$ at $R=20h^{-1}$Mpc for the correlation configuration $n=3$). Note also the neat appearance of the characteristic  baryon acoustic peak at the scale $R\sim 25h^{-1}$Mpc  when the correlation length is computed for $n=4$.   We interpret these results as a qualitative indication of the fact that,  at least on the scales explored by our analysis,  the linear biasing parameter  is well approximated in terms of a  scale-free parameter.  

On the right panel of Figure \ref{d2bigboss} we show the redshift dependence of the $\xi_{g,R}$ for a given arbitrary smoothing scale (in this case $R=17h^{-1}$Mpc). 
The constant amplitude of this statistic, together with the fact that the corresponding matter statistics decreases  by no more than $\sim 0.05$ in amplitude over the same redshift  interval ($0.15<z<0.55$, see Figure \ref{sigestimate}),  provide evidences that linear biasing was nearly $\sim 15\%$ stronger at the early epoch $z=0.55$.
 
In Figure \ref{d3bigboss} we show the scaling of the one- and two-point reduced galaxy cumulant moments  of order 3, namely  $S_{3,g,R}$ and $C_{12,g,R}(r)$. 
The slight and systematic decrease of $S_{3,g,R}$ as a function of scales, much less pronounced than that of $S_{3,R}$  (see Figure \ref{betatheo}), is not compatible with biasing being  described by a single constant parameter $b_1$. Since, we have already argued that  $b_1$ is scale independent, this implies  that the biasing function is non linear and the next order biasing coefficient  $b_2$
must show some scale dependency.

Additional information about the clustering of large scale structures can be retrieved from the analysis of the  complementary third-order galaxy statistic, i.e. the reduced correlator $C_{12,g,R}(r=nR)$
on scales $R$ where this indicator is not too noisy. For both the correlation configurations  $n=3$ and $n=4$ this requirement  limits the region of interest to the scales $R \leq 25h^{-1}$Mpc. 
Despite the fact that measurements on different scales are correlated, it appears that both these statstics are fairly independent from  $R$. 
We remark that in the LS limit  the value of $C_{12,g,R}$ is  also independent from the correlation scale $r$. As a consequence, eq.  (\ref{srealspace}) is best  evaluated by adopting  the smallest possible value of $n$, i.e. the one that minimizes the amplitude of the errorbars. 

Our analysis shows  that both the amplitude of $S_{3,g,R}$ and $C_{12,g,R}(r)$  are nearly redshift independent, i.e.  mostly independent from the cosmic epoch at which these statistics  are computed. This property holds in both real and redshift spaces and  mirrors the analogous behavior, predicted by theory,  for the reduced cumulant moment of the matter field (see eqs. (\ref{s3prediction}) and (\ref{c12prediction}).) Because of this we can conclude that, at least up to redshift $z \sim 0.55$,  even the next  order biasing coefficient, i.e. $b_2$,  is weakly sensitive to time in the Horizon simulations.

The WNLPT  predicts that  reduced moments and correlators of the matter field should  display hierarchical properties in real space. 
Figure \ref{d3bigboss} shows that  the scaling  predicted by  eq. (\ref{hs1})  still holds in redshift space, a results originally found by
Lahav et al. 1993 and Hivon et al. 1995 who showed that it holds  even on smaller scales than those analyzed here, i.e. on domains where non-linear effects become important. 
In an analogous way,  Figure  (\ref{d3bigboss})   shows that the mapping between real and redshift space also  preserves the hierarchical properties  of the reduced correlator $C_{12,g}$.  This property is not a characteristic of low orders statistics only.  In Figure \ref{highcor} we present  the estimates of   the galaxy reduced correlators up to order 5 in both real and redshift space.   Observations in the local universe have shown  that the reduced cumulants  $S_{N,g,R}$ of the smoothed galaxy field in redshift space  display  hierarchical clustering properties up to order $N=6$ (e.g. Baugh et al. 2004).  This plot  shows  that also the reduced correlators  $C_{nm,g,R}$ extracted form the Horizon simulation  preserve the hierarchical scaling up to order $5$ in redshift space.  
Note that, as already anticipated in section \S 5, the amplitude of the clustering  is larger 
in redshift space and  that the relative difference with respect to real space estimates increases systematically as a function of the order of the reduced correlators. 

Finally, Figure \ref{highcor}  graphically displays the validity of the  factorization property that we have found in eq. (\ref{faco}), i.e. that not all galaxy reduced correlators contain original   information.   We stress that even if the factorization property $C_{22,g}=C^{2}_{12,g}$ was shown to hold analytically only in real space, simulations now show that it holds also in redshift distorted space.
 
We conclude by commenting on  the precision of the estimates. The relative error in the estimation of reduced correlators increases, as expected,  with the order of the statistics. Moreover, two-point statistics  are recovered with larger uncertainty than the one-point statistics of the same order  since they are estimated using a smaller number of independent cells. Specifically we find that, the larger the correlation scale,  the stronger the sensitivity of the estimates to finite volume effects. On the typical scale $R=20h^{-1}$Mpc the relative error  with which $C_{12,g}$, $C_{13,g}$, $C_{14,g}$, $C_{22,g}$ and $C_{23,g}$ are  recovered is $4$, $5$, $10$, $15$ and  $33\%$ respectively.
This can be compared to the precision with which the equal-order  reduced moments $S_N$ have been estimated on the same scale, that is $0.7$, $2$ and $10\%$, for the order $3$, $4$ and $5$  respectively.

\subsection{Estimation  of  $\sigma_R(z)$}

In this section we test  the efficiency of the estimator given in eq. (\ref{sigestimz}).  Three major  potential issues, if not properly addressed,  may affect its reliability.  
First, it is imperative to test whether we can safely  apply   WNLPT results in the LS approximation  to compute  the reduced correlators $C_{nm, R}(r)$ in the limit in which 
the cell separation is as low as  $r \sim 3R$, as the  analysis of \citet{BernardeauC} suggests.
Second, we want to verify that peculiar velocity corrections as well as  redshift-space observations do not introduce unexpected 
biases into  our real-space observables.  Finally,  we want to test if  the local Poisson model fairly corrects for the sampling noise in the low counts regime. 

To fulfill these goals, we apply the estimator given in eq. (\ref{sigestimz}) to the simulated LRGs catalogs and gauge the precision with which we can retrieve the real-space amplitude and scaling of $\sigma_R$, that is  both the local normalization and evolution of  the linear matter perturbations embedded in  the $\Lambda$CDM simulations. 

The following argument help us to select the range of scales $R$ that are best suited for applying the formalism to the simulated catalogs. We expect that  the estimator  given in eq. (\ref{sigestimz})     
will work  neatly on sufficiently large scales $R$ (where the WNLPT  and the linear modeling of redshift distortions both apply) and on sufficiently large correlation lengths $r=nR$ (where the LS approximation applies). On the smallest scale where the method  can be theoretically  applied, i.e. $R=10h^{-1}$Mpc,  the amplitude of $\sigma_R$, which is of order $\sim 0.4$  at the  average redshift of the sample,   is completely dominated by shot noise which is  of the order $\sim 1$.  The signal  becomes dominant with respect to discrete sampling corrections  as soon as $R$ is greater than  $\sim 15h^{-1}$Mpc.  Moreover, below this last scale,  a small, but statistically significant imprecision  arises in  assuming $\tau_{g,R}=\tau^{z}_{g,R}$,  as shown in 
Figure \ref{taugrz}. On the opposite end, the largest scale $R$ accessible is set by the geometry of the survey and the requirement of  sampling the correlation length $r=nR$ 
with sufficient statistical power.   We find that the relative error in our estimate of  $C_{12, g,R}(r)$  becomes larger  than  $\sim 10\%$ (see Figure \ref{d3bigboss}) 
for scales $R>22h^{-1}$Mpc when $n=3$ and for scales $R>18 h^{-1}$Mpc when $n=4$. 

\begin{figure}
   \centering
   \includegraphics[width=10cm]{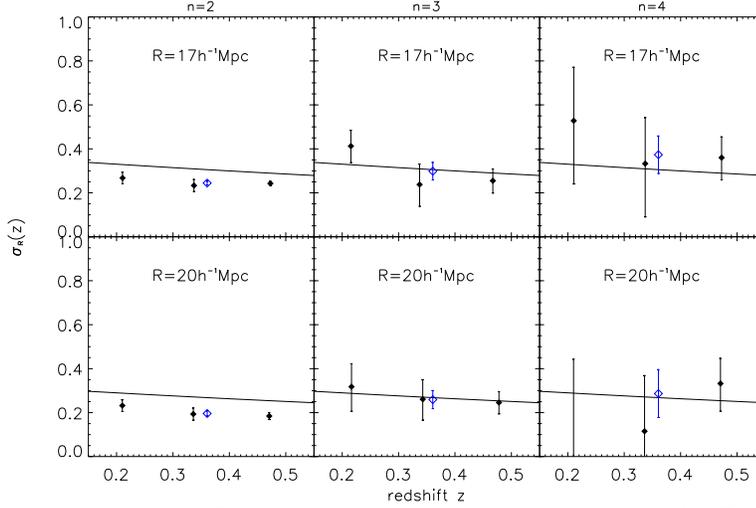}
   \caption{\small The real-space $rms$ of the matter fluctuation field is shown as a function of redshift. Black filled points represent estimates obtained by implementing eq. (\ref{sigestimz}) to the LRGs mock catalogs. Solid lines show the theoretical prediction  obtained by inserting all the physical and  cosmological parameters of the Horizon simulation  into  eq.  (\ref{teosigma}). Count-in-cell and cell correlation analyses  have been performed by assuming spherical cells of radius $R$ separated by correlation lenght $r=nR$. We present results obtained for $n=2$ (left panels), $n=3$ (central panels) and $n=4$ (right panels) using  two typical cell sizes $R=17 h^{-1}$ Mpc (upper panels)  and $R=20h^{-1}$Mpc (lower panels). Each point is the average of the results obtained from eight independent full-sky LRGs catalogs containing nearly 3.8 million galaxies each. Errorbars are estimated  as the standard  error of the mean. At the mean redshift of the catalogs  ($z=0.34$) we also display the average estimate of $\sigma_R$ obtained in the whole survey volume (unfilled blue points).}             
   \label{sigmascal}
\end{figure}

In Figure \ref{sigmascal} we plot the estimates of $\sigma_R$   based on LRGs mock catalogs. We recover the $rms$ of the linear matter fluctuation field
on  two smoothing scales ($R=17$ and $20h^{-1} $Mpc)   and for three different correlation lengths  ($n=2$, $n=3$ and $n=4$). 
By contrasting  our measurements  against the theoretical predictions obtained by inserting into  eq.  (\ref{teosigma}) the parameters used in the Horizon simulations, 
we find that our reconstruction scheme fails when the correlation length is as low as  $n=2$.  This was expected since  \citet{BernardeauC}  already showed that  the LS approximation does not hold on such small correlation scales.  
 Effectively, when we probe larger scales ($n=3$),   the reconstruction  becomes significantly more accurate (central panel of Figure \ref{sigmascal}), with the estimates of 
$\sigma_R$ at  the average redshift of the catalog ($z=0.34$) being affected by a relative error of of $13\%$ and $15\%$ on the scales $R=17$ and $20 h^{-1}$Mpc  respectively. 
For  $n=4$ errorbars become too large for the estimates to be also precise. Additionally,   we  remark that our estimates seem to 
slightly overestimate the value of $\sigma_R$ on both the scale analyzed. This is also  confirmed by the analysis of an independent mock catalog, i.e. the Las Damas simulation \citep{mcb}.
As stressed in section \S 2.3, this is due to the increasing inaccuracy in the theoretical prediction of the amplitude of the reduced correlators $C_{12,R}$ on  correlation lengths that approaches the scale where $\xi_R$ crosses zero.

We re-emphasize that these estimates are totally independent from any assumption about shape and normalization of the 
power spectrum of linear matter fluctuations.  They are also independent from the amplitude of the present day normalization of the 
 Hubble parameter $H_0$. Our estimates, however, do depend on the set of cosmological parameters  ($\Omega_M,  \Omega_{\Lambda}$) that we have  
used to  assign galaxies to cells (i.e.  to smooth the galaxy distribution),  and to subtract the effect of redshift space distortions (i.e. to  evaluate  the growth rate function $f$).

\subsection{Estimation of  the local value of $\sigma_8$}

As we have already  discussed,  the scale $R=8h^{-1}$Mpc falls outside the range of applicability of the test. Nonetheless, we can
extract information about the  value of $\sigma_8$ at redshift $z \sim 0$ ($\sigma_8(0)$) from measurements of  $\sigma_R$ on larger scales. 
We do this by fitting eq. (\ref{teosigma}) to data. The price to pay is that  the recovered value will depend  on the
adopted power spectrum model and on  the set of parameters on which the power spectrum itself  depends, i.e. the reduced Hubble constant $h$, 
the primordial spectral index  $n_s$, and   the reduced density of baryons ($\Omega_b$).   This approach, however offer some advantage:  
if  the relevant cosmological parameters  $\Omega_m$, $\Omega_{\Lambda}$ and $h$ are considered known from  independent probes,  then 
 we can extract information  about the purely gravitational sector of the theory.
 
\begin{figure}
   \centering
   \includegraphics[width=10cm]{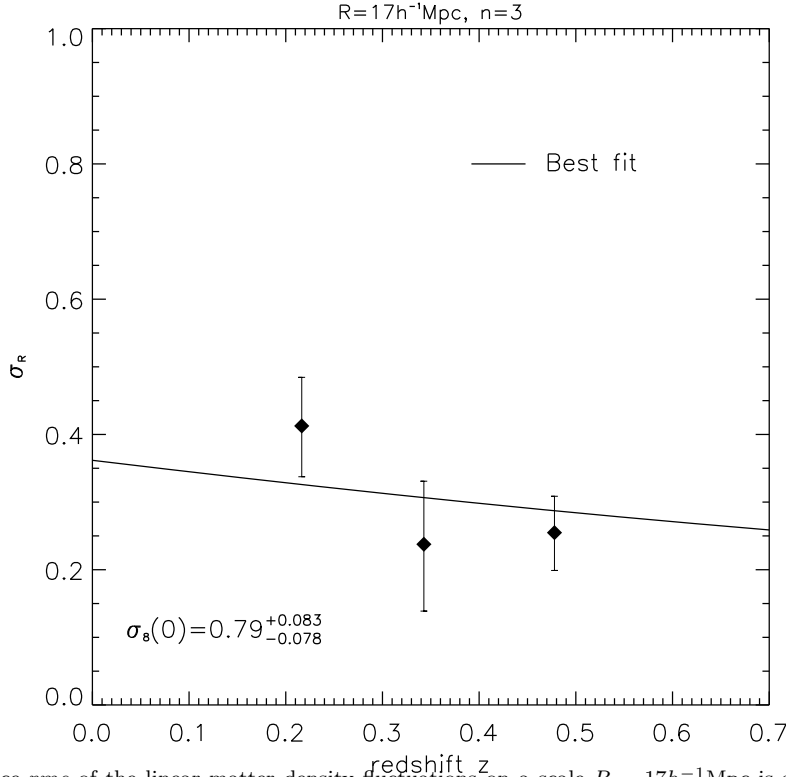}
   \caption{\small  The real-space $rms$ of the linear matter density fluctuations on a scale $R=17h^{-1}$Mpc is shown as a function of redshift. The same data as in the upper central panel of  Figure \ref{sigmascal} are used.
 The solid curve represents the  best fitting theoretical model for the linear scaling of $\sigma_{17}$  (cfr. eq. (\ref{teosigma}) obtained after fixing the amplitude  of $\Omega_b$, $n_s$ and $h$
 to the values of the Horizon simulation.  The corresponding best fitting value of $\sigma_8(0)$  is shown in the inset together with its standard deviation.}               
   \label{sigestimate}
\end{figure}

The way we proceed is as follows: we assume standard gravity as described by GR, we frame our analysis in the linear regime, i.e. we adopt the phenomenological description of the matter power spectrum  given by  \citet{EH}, as well as  the time evolutionary model for $\sigma_R$ given in  eq.  (\ref{teosigma}), and  we look for the  the best fitting parameter $n_s$, $\Omega_b$ and $\sigma_8(0)$ that minimize the statistical distance between our measurements (cfr. eq. (\ref{sigestimz})) and theoretical predictions (cfr. eq. (\ref{teosigma})). The outcome of this approach is displayed in  Figure \ref{sigestimate}.

\begin{figure}
   \centering   
   \includegraphics[width=5.7cm]{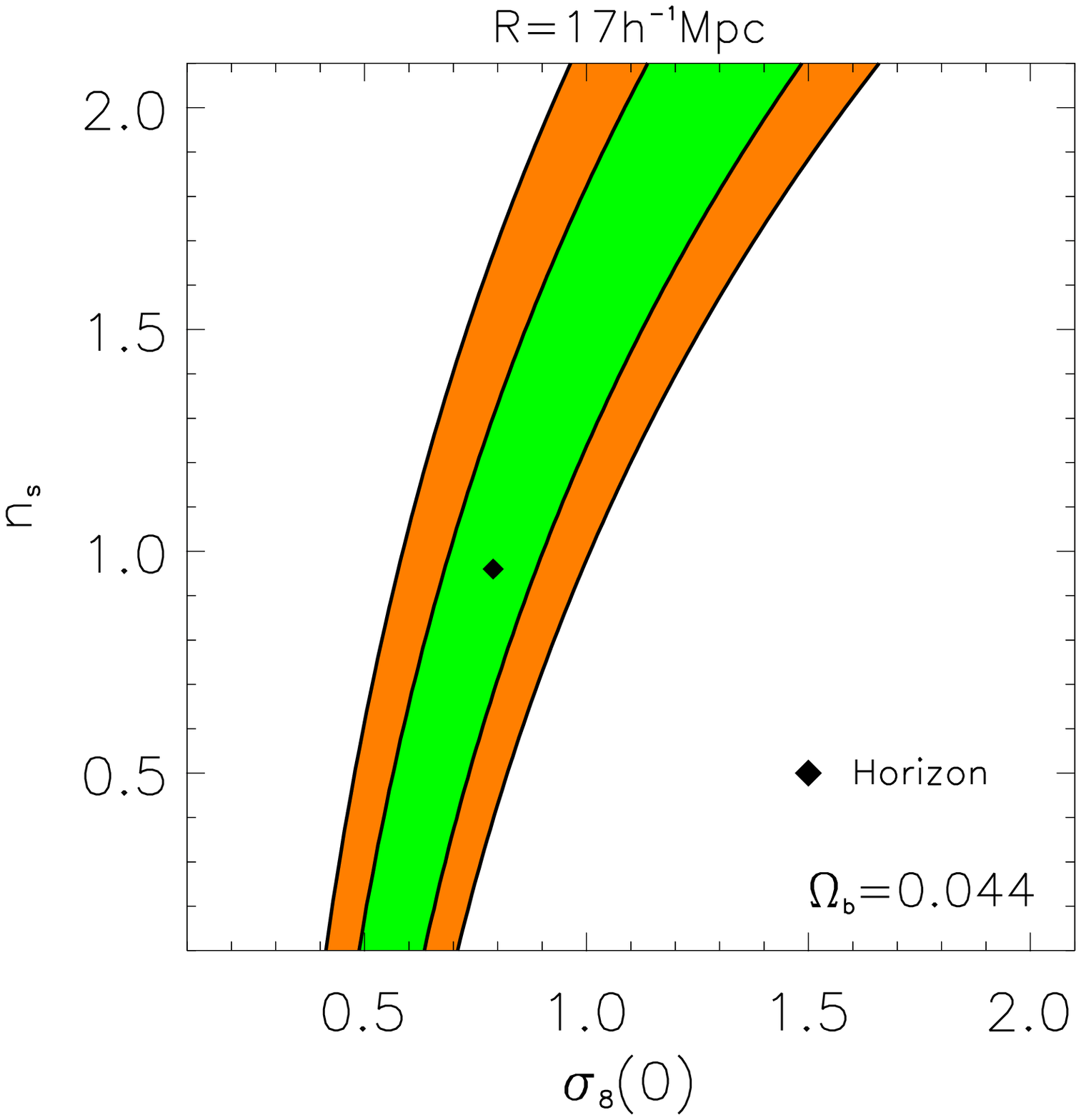}
   \includegraphics[width=5.7cm]{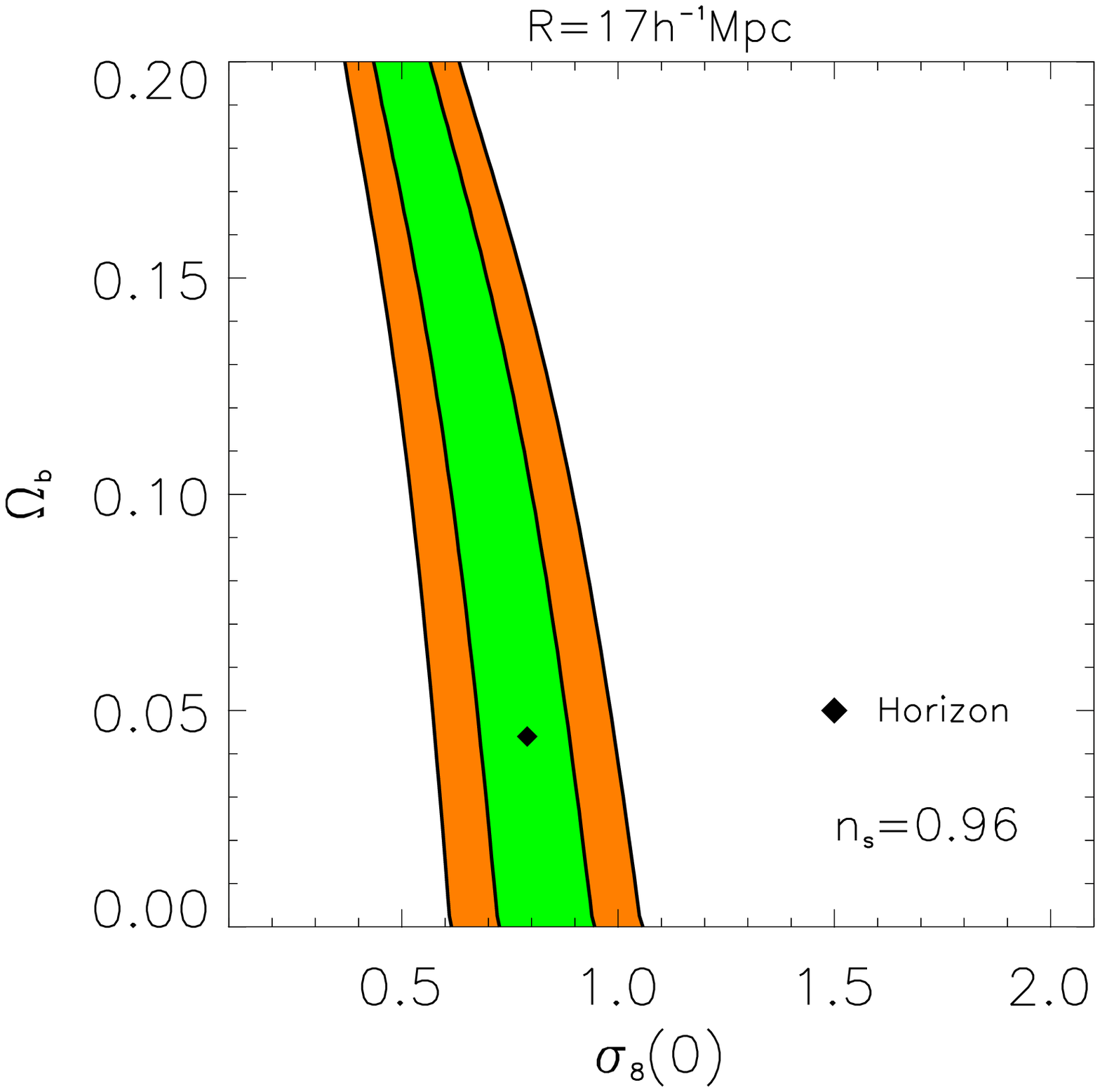}
   \includegraphics[width=5.7cm]{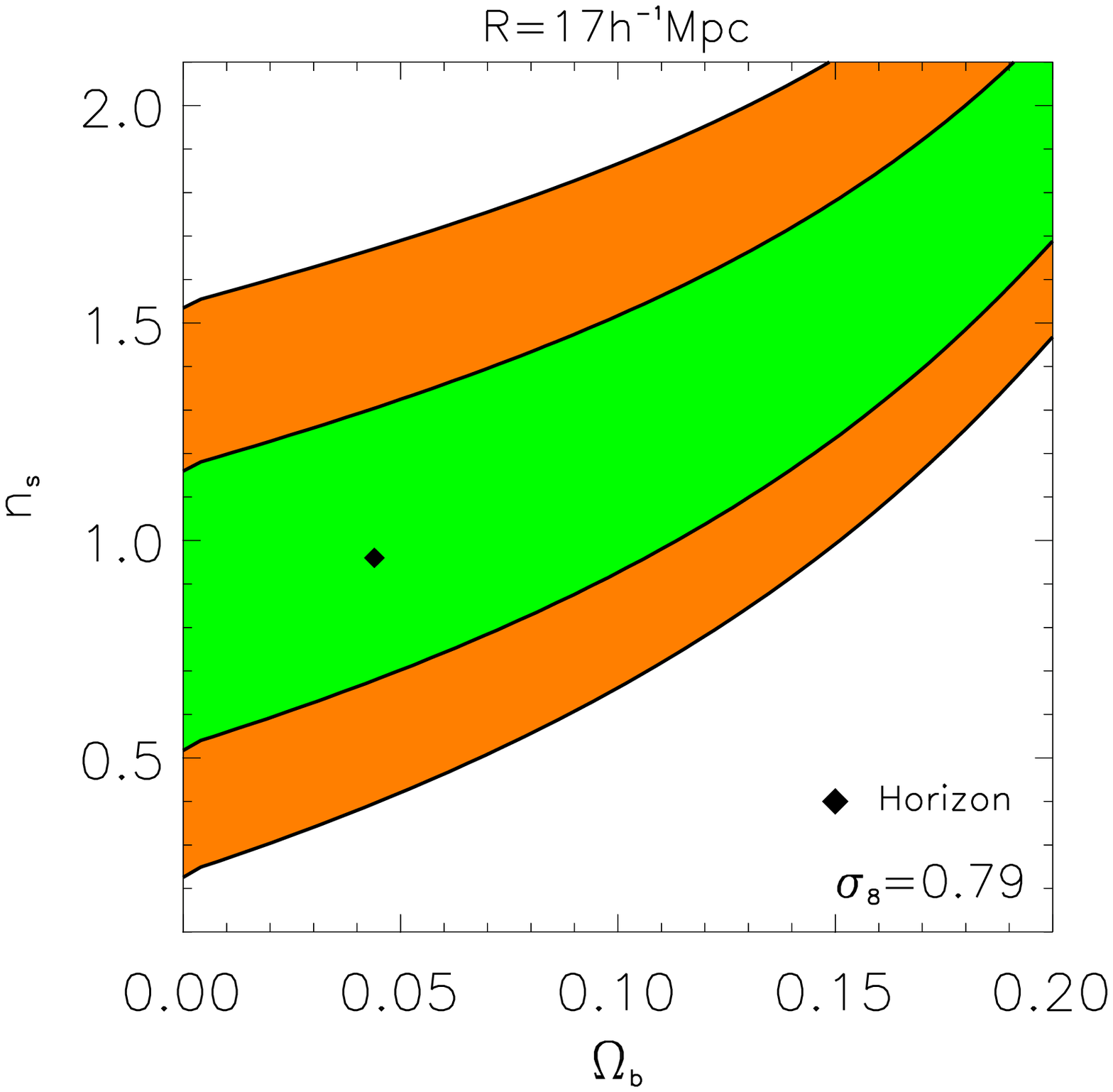}
   \caption{\small   The likelihood contours obtained from the joint estimation of ($\sigma_8(0),n_s$), ($\sigma_8(0),\Omega_b$) and ($\Omega_b,n_s$).
    The analysis have been performed calculating reduced correlators on a scale $R=17h^{-1}$Mpc and at the correlation lenght $n=3$. Isocontours of differently colored region 
    corresponds to $\mathcal{L}_{min}+2.30$ and $\mathcal{L}_{min}+4.61$ where $\mathcal{L}$  is proportional to the log of the likelihood of the data and it is here assumed to be affine 
    to the $\chi^2$ statistic.. 
    On each panel the filled points represent the  fiducial value  of  the Horizon simulation.}             
   \label{joint_ns_sig}
\end{figure}

If we assume that $n_s$ and  $\Omega_b$ are fixed to the simulation's values  $(n_s=0.96, \Omega_b=0.044)$,   we obtain $\sigma_8(0)=0.79\pm 0.08$.  This best fitting value   is  in perfect agreement with the simulated one ($\sigma_8=0.79$). 
Results obtained by  performing a joint two-parameter analysis  (after fixing the third parameter  to the simulated value)  are shown in Figure \ref{joint_ns_sig}. 
The  left and central panel of this figure  reveal that $\sigma_8(0)$ is only marginally degenerate with respect to both $n_s$ and $\Omega_b$, a fact that 
highlights  the fundamental inefficiency of our probe  in constraining the  values of both these parameters.  This conclusion is reinforced in the right panel of the same figure, which displays 
a strong degeneracy between $n_s$ and $\Omega_b$ together with a  loosely constrained confidence region in the corresponding parameters plane.
Luckily, this means that the uncertainties with which  both these parameters are  
estimated using more sensible probes, do not critically  affect the precision with which our method constrain the amplitude of  $\sigma_8(0)$.

Despite the fact that our analysis was performed by slicing the survey volume in independent redshift shells, the limited redshift interval explored  allows us to fix only  the local amplitude of the linear matter fluctuation field.  
In a future work we will show that by implementing  the method with deeper mock catalogs simulating the region of space that will be surveyed spectroscopically by surveys like BigBOSS and EUCLID, 
one can further aim at constraining the time  evolution of $\sigma_R$. 
Data on a  larger redshift interval will allow to  constrain  not only $\sigma_8(0)$,  but also the growth index $\gamma$ in terms of  which the growth rate is  usually 
parameterized  ($f=\Omega_m^{\gamma}(z)$ \cite{pee80}).   This will allow to reject possible alternative description  of gravity, or, in turn the standard model of gravitation itself.

\subsection{Consistency tests}

We have shown that, by using simulations,  it is pretty straightforward  to assess whether the proposed  measuring strategy is able to recover the underlying  value of $\sigma_R$.
What if,  instead,   a real redshift survey is considered? Are there  specific physical criteria or statistical indicators  that  guarantee us that the recovered value of $\sigma_R$ is the {\it true} one?
In other terms we want to shift our attention from the precision of the estimates to their accuracy. 
Apart from the unbiasedness of the WNLPT results in the LS limit,   our test strategy strongly relies on  assuming  that the correct set of cosmological parameters 
($\Omega_M$, $\Omega_{\Lambda}$) has been used in the analysis. As a consequence, any imprecision in the measurements of the reduced cosmic densities translates into a biased estimate of $\sigma_R$. In this section,  we design a diagnostic scheme to test  the coherence of our results and,  at the same time, the soundness of the adopted  values for  $\Omega_M$ and $\Omega_{\Lambda}$.

In the approach developed in this paper,  the linear biasing parameter in real space is 
directly estimated from redshift space observables of  intrinsic third-order nature using the estimator

\begin{equation}
b_{1,R}=\frac{\alpha^{z}_{g,R}}{\tau^{z}_{g,R}}.
\label{nostro}
\end{equation}
As discussed in Section \S 5,  this estimator  has the remarkable 
property of being approximately  independent from cosmology.

 Now  let's  define two new estimators of the real space linear biasing parameter as
 $\mathring{b}_{1,R}=\sigma_{g,R}/\sigma_{R}$ and {\bf $\tilde b_{1,R}=\sqrt{\xi_{g,R}/\xi_{R}}$}, 
where now we exploit, as it is usual,  second order statistics. Both $\sigma_{g,R}$ and $\xi_{g,R}$ are quantities not directly measurable,  nonetheless we
can  recast the expressions of the real space  linear biasing parameters in terms of redshift space observables. By  adopting the Kaiser model for linear motions we  obtain

\begin{eqnarray}
\displaystyle \mathring{b}_{1,R}&=&-\frac{f}{3}+\sqrt{\left(\frac{\sigma_{g,R}^z}{\sigma_R}\right)^2-\frac{4}{45}f^2}   \label{biassrspace} \\
\displaystyle\tilde b_{1,R}&=&-\frac{f}{3}+\sqrt{\frac{\xi_{g,R}^z}{\xi_R}-\frac{4}{45}f^2}.
\label{biasxrspace}
\end{eqnarray}

\begin{figure}
   \centering   
   \includegraphics[width=9.5cm]{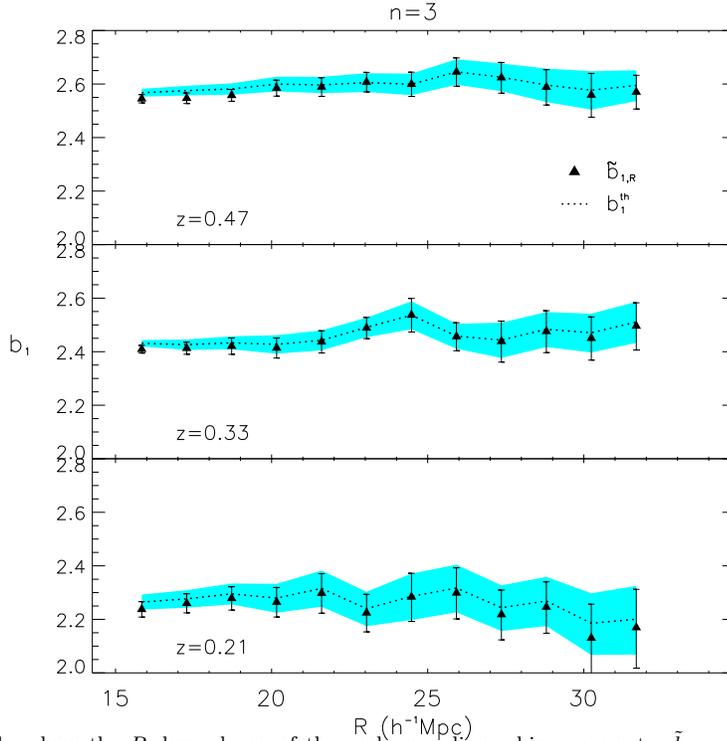}     
   \caption{\small  Black triangles show the  $R$ dependence of the real-space linear bias parameter $\tilde b_{1,R}$  estimated from redshift-space mock catalogs using eq. (\ref{biasxrspace}). 
    Measurements are performed  using the correlation length $n=3$ and in three different redshift bins. These measurements are compared to the 
    real-s[ace  linear bias parameter  $b_1^{th}$ extracted  from real-space mock catalogs using the estimator $\sqrt{\xi_{g,R}/\xi_{R}}$ (dotted line). 
     The shaded  area represents the region of $1$-$\sigma$ uncertainty.} 
   \label{kaiser}
\end{figure}

Before proceeding further, note that in this paper we have assumed that the Kaiser linear modeling of redshift space 
distortions applies on the scales we are interested in. We  can now verify this statement by using the LRGs synthetic catalogs.
To this purpose, and without loss of generality, be  $\tilde b_{1,R} $ the value of the real-space linear bias parameter estimated from  the mock catalogs as $\sqrt{\xi_{g,R}/\xi_{R}}$.
Let's refer to this estimate as to the {\it true} value of the linear-bias parameter and let's  label it as  $b_1^{th}$. In figure \ref{kaiser} we compare 
 these measurements against the estimates of the real-space linear bias inferred using  eq. (\ref{biasxrspace}) in three different redshift intervals and for various 
 smoothing scales $R$.  
 One can see that for $n=3$, the estimator (\ref{biasxrspace}) fairly recovers the real-space value of the  linear bias parameter. This result lends 
 support to   the hypothesis that, at least in the Horizon simulations, and 
 over the range of $R$ scales where we trust the correlator's theory, the distortions induced by large scale  peculiar motions  are accurately described by the Kaiser model.

Physically,  we expect that, whatever is the chosen scale $R$, if $\sigma_8(0)$ has been consistently determined, then eqs.  (\ref{nostro}),  (\ref{biassrspace}) and (\ref{biasxrspace}) all  give the same numerical result.  Since the linear biasing estimators given in eqs. (\ref{biassrspace}) and (\ref{biasxrspace}) depends on the chosen background cosmology,  the correct set of cosmological parameters 
is  thus the very one that makes all the three different linear  biasing definitions converge to the same numerical value on all the scale $R$.
In a different paper (Bel \& Marinoni in prep) we show how this observation can be exploited to guess the background cosmology.  In this paper,   we use this property  to gauge 
the consistency  of our measurements of $\sigma_R$,  
that is to verify that all the different  estimators of the linear biasing parameters  match only when the analysis is carried out in the proper cosmological background.
    
\begin{figure}
   \centering
   \includegraphics[height=7.cm]{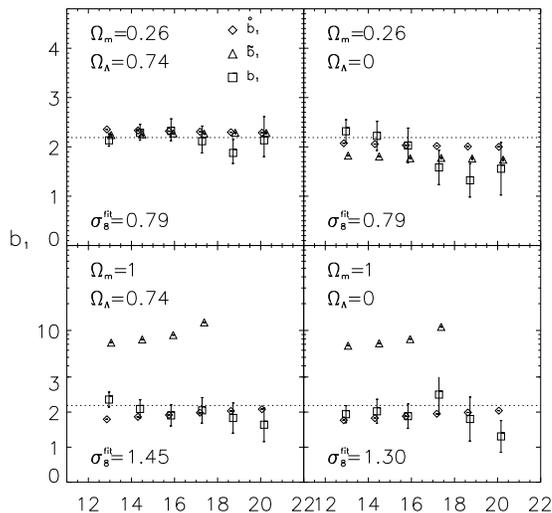}
   \caption{\small  Comparison between the values of the linear bias parameter on various scales $R$. Measurements are obtained  using $3$ different estimators (cfr. eqs. (\ref{nostro}), (\ref{biassrspace}), and (\ref{biasxrspace}))   and by analyzing the same redshift data in  4 different cosmological backgrounds as specified in the panels. 
   $\mathring{b}_{1,R}$ and $\tilde b_{1,R}$ 
   are estimated on the basis of the best fitting value of $\sigma_8(0)$  indicated in each panel. This last value  has been obtained from measurements  of $\sigma_R$ on a scale $R=16h^{-1}$Mpc  
   and by  assuming, arbitrarily, that the linear power spectrum parameters are subject to the following constraints:   $h=0.72$ and  $\Omega_b/\Omega_m=0.17$ for all  the cosmological models.   
    Note that in the lower panel,  $\tilde b_{1,R}$ data are missing on a scales $R=(19,20)h^{-1}$Mpc. This is due to the fact that  the argument of  $\sqrt{\xi_{g,R}/\xi_{R}}$  is negative, i.e. in these
    extreme cosmological models,  the predicted correlation function of the matter density field becomes negative on scales where the corresponding galaxy statistic is still positive.
    The horizontal dotted line shows the average of the measurements obtained with our estimators $b_{1,R}$ in the {\it true} cosmological model (i.e. the $\Lambda$CDM model of the
     Horizon  simulation) and it is reproduced identically in all the panels. This line graphically helps to highlight  the relative insensitivity of our estimator
      $b_{1,R}$ to the choice of the parameters  ($\Omega_M, \Omega_{\Lambda}$). Note  how different estimates  converge to the same value only when the analysis is performed in the {\it true} cosmological model.     } 
   \label{biastest}
\end{figure}

 In Figure   \ref{biastest} we  perform this test  and  show what  pathological  features  do show up when  the analysis relies on an improperly chose set of values 
 ($\Omega_M, \Omega_{\Lambda}$).    The algorithm goes as follows: we estimate $\sigma_R$ on a given arbitrary scale $R$ (here we chose $R=16h^{-1}$Mpc), 
 and in four different cosmologies (indicated in each panel of Figure \ref{biastest}). We then deduce  the value of  $\sigma_8(0)$ in each of 
 these four scenarios.  In doing this, we implicitly assume  that the specific cosmology adopted in order to  measure $\sigma_{R=16}$ is the correct one and that, as a consequence, 
 the value of $\sigma_8(0)$ inferred using any  other scale $R$ {\it is identically the same}. We then plug in this value of $\sigma_8(0)$  into eqs. eqs. (\ref{biassrspace}) and (\ref{biasxrspace}) and compare the results with those obtained via the estimator  $b_{1,R}$, i.e.  we contrast them against a measurement of the linear biasing parameter  that is weakly sensitive to  cosmology.   
 
 If we analyze the LRGs  mock catalogs  by assuming the simulated  set of cosmological parameters ($\Omega_M=0.26, \Omega_{\Lambda}=0.74$, see upper left panel of 
 Figure \ref{biastest}) then the estimates  of $\mathring{b}_{1,R}$, $\tilde b_{1,R}$ and $b_{1,R}$ are consistent between themselves  on all the scales $R$.  
 On the contrary,  if we  process data by incorrectly assuming  a low density (open)  background model (upper right panel in Figure \ref{biastest}), the different estimations of the linear biasing 
 parameter  are not anymore in agreement, even if  the estimated value  of $\sigma_8(0)$ has not  changed (note that the linear power spectrum at present epoch is independent from the amplitude of the  cosmological constant and thus insensitive to its variation). In particular, 
 $\mathring{b}_{1,R}$ identically coincides,  by definition,  with the measure obtained using our  estimator $b_{1,R}$ only  for $R=16h^{-1}$Mpc , but the estimates deviate on all the other scales.
This  discrepancy is due to the fact that, in a ``wrong" cosmology,  the  extrapolated value of $\sigma_8(0)$  may not  be  independent from the scale $R$ on which $\sigma_R$ is measured.  Or said differently, the equation  $\mathring{b}_{1,R}(\sigma_8(0)) = b_{1,R}$ might not have  a unique solution $\sigma_8(0)$ for all the scales $R$.  

The  effect previously discussed is mostly the consequence of adopting the wrong amplitude for the cosmological constant.  It 
 is, however, less pronounced than the discrepancy between the measurements of $\mathring{b}_{1,R}$ and $\tilde b_{1,R}$ that  arises when the value of $\Omega_m$ is 
 poorly guessed. The effects of a wrong choice of the matter density parameter  are  presented  in the lower panels 
 of Figure \ref{biastest}.  The observed  large inconsistency   arises  essentially from  the fact that the zero order spherical 
 Bessel function appearing inside the integral in eq. (\ref{gg}) filters in different portions of the signal (i.e. of  $\Delta^{2}_{L}$) if  the characteristic parameters of the linear power spectrum are changed. In other terms, if we consider eq. (\ref{gg}) and spuriously overestimate $\Omega_m$, the predicted suppression of  power  on a scales  $r$ is larger than the variation actually seen in the data. 
 
 We remark that our estimator of the linear biasing parameter, relying on third-order statistical indicators,  is affected by  errorbars that are larger than those associated to the classical estimators $\mathring{b}_{1,R}$  and $\tilde{b}_{1,R}$. This imprecision is largely compensated by the fact that our estimator does not depend on any assumption about the nature of the dark matter (i.e. the specific form of the matter power spectrum) and it is almost insensitive also to its abundance ($\Omega_M$).  Therefore, by contrasting different estimations of the linear biasing parameter, using the diagnostic diagram of Figure \ref{biastest},  we can deduce if the scaling of $\sigma_R$ was inferred in the appropriate cosmological model.

\section*{Conclusions}

A key goal that seems to  fall within the technical reach of future cosmological experiments is   
to rule out (or in) eventual infrared modifications of the standard theory of gravity,  as possibly manifested  by  the  unexpected growth of  cosmic mass in the weakly non-linear,  low-curvature, high-redshift regimes.  To fulfill this  task,  it is mandatory to devise sensible observables of the large scale structure of the universe. 
In this spirit, this paper focuses on the  potential  of  clustering indicators that have been rarely explored  in literature, that is the high-order reduced correlators $C_{nm}$ of the 3D  mass  overdensity  field. 

To fully exploit the richness of information contained in this  two-point statistics, whose amplitude is analytically predicted by the weakly non-linear perturbation theory in the large separation limit approximation,  we have derived the  expressions of the reduced correlators of the smoothed galaxy density  field ($C_{nm,g}$) up to order $5$.  We have found that  they preserve both  the  hierarchical scaling  and the factorization properties  of the matter reduced correlators. 

Building upon these results we have  worked  out the explicit expressions for the bias coefficients up to order $4$  and a  new estimator to measure the $rms$ of the linear matter fluctuations on a scale $R$ directly from galaxy redshift surveys.  The central
 result of this paper, namely the estimator  given in eq. (\ref{sigestimz}),  has been tested using artificial galaxy catalogs and shown  to recover fairly well the `hidden' simulated value of $\sigma_R$.

Despite the fact that very large survey volume are needed to make the estimation of these observables accurate enough for 
cosmological purposes, the merit of this approach are  evident: 
{\it a)}
linear biasing is not a parameter that one needs to marginalize over,  but a physical parameter that can be estimated in a totally  independently  way  from 
any assumption about the structure of the linear power spectrum and the  value of cosmological parameters. {\it b)}
the real space  linear biasing parameter can be measured directly using 
redshift space observables. {\it c)} The scaling of $\sigma_R$ can be inferred directly without imposing any {\it a-priori} constraint on the eventual non-linear and scale dependent nature of the bias function. {\it d})  The correlator formalism allows also for a self-consistent test of the coherence of the results obtained concerning the  scaling of  $\sigma_R$, a 
step further in  the direction of making cosmological results  not only precise but also accurate. 

In this paper we have analyzed  local simulations with the aim of testing  principles and theoretical ingredients  on which the proposed strategy relies.  
Work is already in progress to apply the formalism  to the SDSS-dr7 data and to extract the local value of linear matter fluctuations on sensible scales $R$. 
The good news is that the next decade holds even greater prospects for growth of the red-shifts data base. 
Therefore, we also plan to implement the algorithm  to mock catalogs simulating future large 3D surveys such as BigBOSS and EUCLID and forecast 
up to what order  and precision the bias coefficients $b_i$ can be estimated, as well as, the figure of merit achievable on $\sigma_8$ 
and on the  gravitational growth index $\gamma$. 

From the theoretical side, valuable insights are expected from  a reverse engineering on the proposed test. 
For example, we show in a different paper (Bel \& Marinoni in prep.) that  if the gravitational model is {\it a-priori} known then   
the formalism offers the possibility of  narrowing in on the value  of fundamental cosmological parameters such as $\Omega_m$ and $\Omega_{\Lambda}$. 
Also   further  work is needed to understand how higher order real-space biasing parameters can be effectively retrieved 
from redshift space reduced correlators. Finally, interesting possibilities will open up if  WNLPT  predictions could be extended into the small separation  limit  
where much more statistical power is locked.

CM is grateful for support from  specific project funding of the {\it Institut Universitaire de France}. 
We thank the anonymous referee for his comments and suggestions.
We acknowledge useful discussions with  E. Branchini,  E. Gazta\~naga, L. Guzzo, L. Moscardini.

\appendix



\section[]{Appendix A: higher order galaxy two-point cumulant moments}

Listed here are the amplitudes of the two-point galaxy cumulant moments. Correlators  of the galaxy distribution have been computed up to order 5.

\begin {eqnarray}
 \kappa_{g, 11} & = &  {b_{{1}}}^{2}\kappa_{11}+{b_{{1}}}^{2}\big ( c_{{3}}+C_{{12}}c_{{2}} \big ) \kappa_{11}\,\kappa_{{2}}+1/2\,{b_{{1}}}^{2}{c_{{2}}}^{2}{\kappa_{11}}^{2} \nonumber \\ 
 \kappa_{g, 12} & =& {b_{{1}}}^{3}\big ( C_{{12}}+2\,c_{{2}} \big ) \kappa_{11}\,\kappa_{{2}}+{b_{{1}}}^{3} \big ( 5/2\,c_{{3}}C_{{12}}+2\,{c_{{2}}}^{2}C_{{12}}+3\,c_{{3}}c_{{2}}+c_{{4}}+1/2\,c_{{2}}C_{{22}}+{c_{{2}}}^{2}S_{{3}}+  c_{{3}}S_{{3}}+c_{{2}}C_{{13}} \big ) \kappa_{11}\,{\kappa_{{2}}}^{2} \nonumber \\
 &  & +{b_{{1}}}^{3} \big ( {c_{{2}}}^{3}+1/2\,c_{{4}}+3\,{c_{{2}}}^{2}C_{{12}}+2\,c_{{3}}c_{{2}}+c_{{3}}C_{{12}} \big ) {\kappa_{11}}^{2}\kappa_{{2}}+{b_{{1}}}^{3}c_{{3}}c_{{2}}{\kappa_{11}}^{3} \nonumber \\ 
\kappa_{g, 13} & =& {b_{{1}}}^{4}\big ( 6\,{c_{{2}}}^{2}+C_{{13}}+3\,c_{{2}}S_{{3}}+3\,c_{{3}}+6\,c_{{2}}C_{{12}} \big ) \kappa_{11}\,{\kappa_{{2}}}^{2}+{b_{{1}}}^{4} 
\big ( 3\,C_{{12}}c_{{2}}+6\,{c_{{2}}}^{2} \big ) {\kappa_{11}}^{2}\kappa_{{2}}+{b_{{1}}}^{4}c_{{3}}{\kappa_{11}}^{3}+  {b_{{1}}}^{4} \big ( 3/2\,c_{{2}}C_{{14}}\nonumber \\
& & +9/2\,C_{{12}}c_{{3}}S_{{3}}+12\,{c_{{2}}}^{2}c_{{3}}+6\,{c_{{2}}}^{3}S_{{3}}+{\frac{45}{2}}\,c_{{2}}c_{{3}}C_{{12}}+9\,c_{{2}}c_{{4}}+3/2\,c_{{3}}S_{{4}}+ 9/2\,c_{{4}}S_{{3}}+3/2\,c_{{5}}+15/2\,{c_{{3}}}^{2}+3\,{c_{{2}}}^{2}S_{{4}} \nonumber \\
 & & +3\,{c_{{2}}}^{2} C_{{22}} +15/2\,{c_{{2}}}^{2}C_{{12}}S_{{3}}+9/2\,c_{{4}}C_{{12}}+ 6\,{c_{{2}}}^{2}C_{{13}}+{\frac {45}{2}}\,c_{{3}}S_{{3}}c_{{2}}+5\,c_{{3}}C_{{13}}+6\,{c_{{2}}}^{3}C_{{12}}+1/2\,c_{{2}}C_{{23}} \big ) \kappa_{11}\,{\kappa_{{2}}}^{3} \nonumber \\
 & & +{b_{{1}}}^{4}  \big ( 9/2\,{c_{{2}}}^{2}{C_{{12}}}^{2}+ 6\,{c_{{2}}}^{2}C_{{13}}+3/2\,c_{{3}}C_{{22}}+{\frac {39}{2}}\,c_{{2}}c_{{3}}C_{{12}}+3/2\,c_{{4}}C_{{12}}+9/2\,c_{{3}}S_{{3}}c_{{2}}+3\,{c_{{2}}}^{4}+18\,{c_{{2}}}^{3}C_{{12}}+  3/2\,c_{{3}}{C_{{12}}}^{2} \nonumber \\ 
 & & +15/2\,c_{{2}}c_{{4}}+18\,{c_{{2}}}^{2}c_{{3}}+6\,{c_{{2}}}^{3}S_{{3}} \big ) {\kappa_{11}}^{2}{\kappa_{{2}}}^{2} +{b_{{1}}}^{4} \big ( 9/2\,{c_{{3}}}^{2}+1/2\,c_{{5}}+3/2\,c_{{4}}C_{{12}}+  9\,c_{{2}}c_{{3}}C_{{12}}+6\,{c_{{2}}}^{2}c_{{3}} \big ) {\kappa_{11}}^{3}\kappa_{{2}} \nonumber \\ 
 & & +3/2\,{b_{{1}}}^{4}c_{{2}}c_{{4}}{\kappa_{11}}^{4}  \nonumber \\ 
 \kappa_{g, 22} & =& {b_{{1}}}^{4} \big ( 4\,c_{{2}}C_{{12}}+4\,{c_{{2}}}^{2}+C_{{22}} \big ) \kappa_{11}\,{\kappa_{{2}}}^{2}+{b_{{1}}}^{4} \big ( 4\,c_{{3}}+8\,C_{{12}}c_{{2}}+4\,{c_{{2}}}^{2}\big ) {\kappa_{11}}^{2}\kappa_{{2}}+4\,{b_{{1}}}^{4}{c_{{2}}}^{2}{\kappa_{11}}^{3}+ {b_{{1}}}^{4} \big ( 2\,{c_{{2}}}^{2}C_{{22}}\nonumber \\
 & & +2\,{c_{{2}}}^{2}C_{{12}}S_{{3}}+4\,c_{{3}}C_{{22}}+4\,{c_{{2}}}^{3}S_{{3}}+2\,c_{{2}}C_{{23}}+12\,c_{{2}}c_{{3}}C_{{12}}+2\,C_{{12}}c_{{3}}S_{{3}}+ 8\,{c_{{2}}}^{2}c_{{3}}+4\,{c_{{2}}}^{2}C_{{13}}+4\,c_{{2}}c_{{4}}+4\,c_{{3}}S_{{3}}c_{{2}} \nonumber \\
 & &  +4\,{c_{{2}}}^{3}C_{{12}}+2\,c_{{4}}C_{{12}} \big ) \kappa_{11}\,{\kappa_{{2}}}^{3}  + {b_{{1}}}^{4} \big ( 2\,{c_{{2}}}^{4}+ 6\,c_{{3}}S_{{3}}c_{{2}}+8\,{c_{{2}}}^{2}C_{{22}}+2\,c_{{4}}S_{{3}}+4\,c_{{3}}{C_{{12}}}^{2}+10\,{c_{{2}}}^{2}{C_{{12}}}^{2}+8\,c_{{4}}C_{{12}}+10\,{c_{{3}}}^{2}\nonumber \\
 & & +2\,c_{{5}}+ 36\,c_{{2}}c_{{3}}C_{{12}}  +8\,{c_{{2}}}^{2}c_{{3}}+4\,{c_{{2}}}^{2}C_{{13}}+ 12\,{c_{{2}}}^{3}C_{{12}}+4\,c_{{3}}C_{{13}}+6\,c_{{2}}c_{{4}} \big ) {\kappa_{11}}^{2}{\kappa_{{2}}}^{2}+{b_{{1}}}^{4}   \big ( 8\,c_{{2}}c_{{4}}+ 16\,{c_{{2}}}^{3}C_{{12}}\nonumber \\ 
 & & +20\,c_{{2}}c_{{3}}C_{{12}}+16\,{c_{{2}}}^{2}c_{{3}} \big ) {\kappa_{11}}^{3}\kappa_{{2}}+{b_{{1}}}^{4}\big (2{c_{{3}}}^{2}+{c_{{2}}}^{4}+4\,{c_{{2}}}^{2}c_{{3}}\big ){\kappa_{11}}^{4} \nonumber \\
\kappa_{g, 14} & =&  {b_{{1}}}^{5} \big ( C_{{14}}+24\,{c_{{2}}}^{3}+12\,c_{{3}}S_{{3}}+4\,c_{{4}}+12\,c_{{2}}C_{{12}}S_{{3}}+36\,{c_{{2}}}^{2}S_{{3}}+12\,c_{{2}}C_{{13}}+36\,{c_{{2}}}^{2}C_{{12}} +   4\,c_{{2}}S_{{4}}+12\,c_{{3}}C_{{12}}\nonumber \\ 
& & +36\,c_{{3}}c_{{2}} \big ) \kappa_{11}\,{\kappa_{{2}}}^{3} +{b_{{1}}}^{5} \big ( 36\,{c_{{2}}}^{2}C_{{12}}+4\,c_{{2}}C_{{13}}+3\,c_{{2}}{C_{{12}}}^{2}+36\,{c_{{2}}}^{3}+ 12\,c_{{3}}c_{{2}}+12\,{c_{{2}}}^{2}S_{{3}} \big ) {\kappa_{11}}^{2}{\kappa_{{2}}}^{2}+{b_{{1}}}^{5} \big ( 6\,c_{{3}}C_{{12}} \nonumber \\
& & +12\,c_{{3}}c_{{2}} \big ) {\kappa_{11}}^{3}\kappa_{{2}}+{b_{{1}}}^{5}c_{{4}}{\kappa_{11}}^{4}+ {b_{{1}}}^{5} \big ( 42\,c_{{2}}c_{{3}}{S_{{3}}}^{2}+ 12\,c_{{3}}S_{{3}}C_{{13}}+8\,c_{{3}}C_{{12}}S_{{4}}+2\,c_{{6}}+90\,{c_{{2}}}^{3}C_{{12}}S_{{3}}+24\,{c_{{2}}}^{2}C_{{13}}S_{{3}}\nonumber \\
 & & +20\,{c_{{2}}}^{2}C_{{12}}S_{{4}} + 6\,{c_{{2}}}^{2}C_{{22}}S_{{3}}+180\,c_{{3}}{c_{{2}}}^{2}C_{{12}}+84\,c_{{2}}c_{{3}}C_{{13}} + 282\,{c_{{2}}}^{2}c_{{3}}S_{{3}}+6\,c_{{2}}c_{{3}} C_{{22}}+62\,c_{{2}}c_{{3}}S_{{4}}+ 74\,c_{{2}}c_{{4}}C_{{12}}\nonumber \\
 & & +114\,c_{{2}}c_{{4}}S_{{3}} +24\,c_{{4}}S_{{3}}C_{{12}}+168\,c_{{2}}c_{{3}}S_{{3}}C_{{12}} +8\,c_{{5}}C_{{12}}+24\,c_{{2}}c_{{5}}+12\,c_{{5}}S_{{3}}+12\,c_{{4}}C_{{13}}+8\,c_{{4}}S_{{4}}+6\,c_{{4}}{S_{{3}}}^{2}  \nonumber \\
 & & +72\,c_{{4}}{c_{{2}}}^{2}+44\,c_{{3}}c_{{4}} +54\,{c_{{3}}}^{2}C_{{12}}+84\,{c_{{3}}}^{2}S_{{3}} +114\,{c_{{3}}}^{2}c_{{2}} +60\,{c_{{2}}}^{3}c_{{3}}+17/2\,c_{{3}}C_{{14}}+2\,c_{{3}}S_{{5}} +1/2\,c_{{2}}C_{{2,4}}+2\,c_{{2}}C_{{1,5}}\nonumber \\
 & & +6\,{c_{{2}}}^{2}S_{{5}}+18\,{c_{{2}}}^{3} C_{{22}}+24\,{c_{{2}}}^{3}{S_{{3}}}^{2}+ 6\,{c_{{2}}}^{2}C_{{23}}+36\,{c_{{2}}}^{3}C_{{13}}+36\,{c_{{2}}}^{4}S_{{3}}+18\,{c_{{2}}}^{2}C_{{14}}+24\,{c_{{2}}}^{4}C_{{12}}+36\,{c_{{2}}}^{3}S_{{4}} \big )\kappa_{11}\,{\kappa_{{2}}}^{4}\nonumber \\
 & & +{b_{{1}}}^{5}\big ( 2\,c_{{3}}C_{{12}}C_{{13}}+3\,c_{{3}}C_{{22}}C_{{12}}+12\,{c_{{2}}}^{5}+20\,{c_{{2}}}^{2}C_{{12}}C_{{13}}+72\,{c_{{2}}}^{3}C_{{12}}S_{{3}}+270\,c_{{3}}{c_{{2}}}^{2}C_{{12}}+ 42\,c_{{2}}c_{{3}}{C_{{12}}}^{2}+32\,c_{{2}}c_{{3}}C_{{13}} \nonumber \\
 & & +156\,{c_{{2}}}^{2}c_{{3}}S_{{3}}+18\,c_{{2}}c_{{3}}C_{{22}} +8\,c_{{2}}c_{{3}}S_{{4}}+54\,c_{{2}}c_{{4}}C_{{12}}+30\,c_{{2}}c_{{4}}S_{{3}}+ 48\,c_{{2}}c_{{3}}S_{{3}}C_{{12}}+8\,c_{{2}}c_{{5}} +2\,c_{{4}}C_{{13}}+3/2\,c_{{4}}{C_{{12}}}^{2}\nonumber \\
 & & 90\,c_{{4}}{c_{{2}}}^{2}+ 6\,c_{{3}}c_{{4}}+12\,{c_{{3}}}^{2}C_{{12}}+48\,{c_{{3}}}^{2}c_{{2}}+144\,{c_{{2}}}^{3}c_{{3}}+2\,c_{{3}}C_{{23}}+ 72\,{c_{{2}}}^{3}C_{{13}}+54\,{c_{{2}}}^{3}{C_{{12}}}^{2}+72\,{c_{{2}}}^{4}S_{{3}}+10\,{c_{{2}}}^{2}C_{{14}}\nonumber \\
 & & +108\,{c_{{2}}}^{4}C_{{12}}+ 18\,{c_{{2}}}^{3}S_{{4}} \big ) {\kappa_{11}}^{2}{\kappa_{{2}}}^{3} +{b_{{1}}}^{5} \big ( 108\,c_{{3}}{c_{{2}}}^{2}C_{{12}}+30\,c_{{2}}c_{{3}}{C_{{12}}}^{2}+20\,c_{{2}}c_{{3}}C_{{13}} +48\,{c_{{3}}}^{2}C_{{12}} +12\,{c_{{3}}}^{2}S_{{3}}\nonumber \\
 & & + 18\,c_{{2}}c_{{4}}C_{{12}}+6\,c_{{2}}c_{{5}}+78\,{c_{{3}}}^{2}c_{{2}}+12\,c_{{3}}c_{{4}}+36\,{c_{{2}}}^{3}c_{{3}}+24\,{c_{{2}}}^{2}c_{{3}}S_{{3}}+3\,c_{{4}}C_{{22}}+3\,c_{{5}}C_{{12}}+ 6\,c_{{4}}{C_{{12}}}^{2} \big ) {\kappa_{11}}^{3}{\kappa_{{2}}}^{2}+\nonumber \\
 & &  {b_{{1}}}^{5} \big ( 2\,c_{{5}}C_{{12}}+1/2\,c_{{6}}+20\,c_{{2}}c_{{4}}C_{{12}} +18\,c_{{4}}{c_{{2}}}^{2}+8\,c_{{3}}c_{{4}} \big ) {\kappa_{11}}^{4}\kappa_{{2}} +2\,{b_{{1}}}^{5}c_{{2}}c_{{5}}{\kappa_{11}}^{5} \nonumber \\
\kappa_{g, 23} & =&  {b_{{1}}}^{5} \big ( C_{{23}}+2\,c_{{2}}C_{{13}}+6\,c_{{2}}C_{{22}}+6\,{c_{{2}}}^{2}S_{{3}}+3\,c_{{3}}C_{{12}}+18\,{c_{{2}}}^{2}C_{{12}}+6\,c_{{3}}c_{{2}}+3\,c_{{2}}C_{{12}}S_{{3}} +12\,{c_{{2}}}^{3}\big ) \kappa_{11}\,{\kappa_{{2}}}^{3}\nonumber \\
& & +{b_{{1}}}^{5} \big ( 6\,c_{{2}}C_{{22}} +18\,c_{{3}}C_{{12}}+12\,c_{{2}}{C_{{12}}}^{2}+3\,c_{{3}}S_{{3}}+6\,c_{{2}}C_{{13}}+6\,{c_{{2}}}^{2}S_{{3}} +18\,{c_{{2}}}^{3}+54\,{c_{{2}}}^{2}C_{{12}}+3\,c_{{4}}+30\,c_{{3}}c_{{2}} \big ) {\kappa_{11}}^{2}{\kappa_{{2}}}^{2}\nonumber \\
& &+ {b_{{1}}}^{5} \big ( 18\,c_{{3}}c_{{2}}+24\,{c_{{2}}}^{3}+2\,c_{{4}}+36\,{c_{{2}}}^{2}C_{{12}}  +6\,c_{{3}}C_{{12}} \big ) {\kappa_{11}}^{3}\kappa_{{2}}+6\,{b_{{1}}}^{5}c_{{2}}c_{{3}}{\kappa_{11}}^{4}+6\,{b_{{1}}}^{5}{c_{{2}}}^{3}{\kappa_{11}}^{4}+{b_{{1}}}^{5} \big ( c_{{2}}C_{{33}}\nonumber \\
& & +3\,c_{{2}}c_{{3}}{S_{{3}}}^{2} +c_{{3}}S_{{3}}C_{{13}}+ 3/2\,c_{{3}}C_{{12}}S_{{4}}+27\,{c_{{2}}}^{3}C_{{12}}S_{{3}}+4\,{c_{{2}}}^{2}C_{{13}}S_{{3}}+3\,{c_{{2}}}^{2}C_{{12}}S_{{4}}+6\,{c_{{2}}}^{2}C_{{22}}S_{{3}}+72\,c_{{3}}{c_{{2}}}^{2}C_{{12}}\nonumber \\
&  &  14\,c_{{2}}c_{{3}}C_{{13}}+57\,{c_{{2}}}^{2}c_{{3}}S_{{3}}+30\,c_{{2}}c_{{3}}C_{{22}}+3\,c_{{2}}c_{{3}}S_{{4}}+24\,c_{{2}}c_{{4}}C_{{12}}+12\,c_{{2}}c_{{4}}S_{{3}}+9/2\,c_{{4}}S_{{3}}C_{{12}}+ 42\,c_{{2}}c_{{3}}S_{{3}}C_{{12}}\nonumber \\
& & +9/2\,c_{{3}}S_{{3}}C_{{22}} +3/2\,c_{{5}}C_{{12}}+3\,c_{{2}}c_{{5}}+c_{{4}}C_{{13}}+9/2\,c_{{4}}C_{{22}}+24\,c_{{4}}{c_{{2}}}^{2}+3\,c_{{3}}c_{{4}}+12\,{c_{{3}}}^{2}C_{{12}}+3\,{c_{{3}}}^{2}S_{{3}}+18\,{c_{{3}}}^{2}c_{{2}}+ \nonumber \\
& &  30\,{c_{{2}}}^{3}c_{{3}}+13/2\,c_{{3}}C_{{23}}+3/2\,c_{{2}}C_{{2,4}}+ 9\,{c_{{2}}}^{3}C_{{22}}+3\,{c_{{2}}}^{3}{S_{{3}}}^{2}+ 13\,{c_{{2}}}^{2}C_{{23}}+18\,{c_{{2}}}^{3}C_{{13}}+18\,{c_{{2}}}^{4}S_{{3}}+3\,{c_{{2}}}^{2}C_{{14}}+\nonumber \\
& & 12\,{c_{{2}}}^{4}C_{{12}}+6\,{c_{{2}}}^{3}S_{{4}}\big ) \kappa_{11}\,{\kappa_{{2}}}^{4}+ {b_{{1}}}^{5} \big ( 12\,c_{{3}}C_{{12}}C_{{13}}+6\,c_{{3}}C_{{22}}C_{{12}}+3/2\,c_{{6}}+6\,{c_{{2}}}^{5}+30\,{c_{{2}}}^{2}C_{{12}}C_{{13}}+42\,{c_{{2}}}^{3}C_{{12}}S_{{3}}\nonumber \\
& & +  279\,c_{{3}}{c_{{2}}}^{2}C_{{12}}+108\,c_{{2}}c_{{3}}{C_{{12}}}^{2}+72\,c_{{2}}c_{{3}}C_{{13}} +84\,{c_{{2}}}^{2}c_{{3}}S_{{3}}+45\,c_{{2}}c_{{3}}C_{{22}}+9\,c_{{2}}c_{{3}}S_{{4}}+ 108\,c_{{2}}c_{{4}}C_{{12}}+36\,c_{{2}}c_{{4}}S_{{3}} \nonumber \\
& & +12\,c_{{4}}S_{{3}}C_{{12}}+72\,c_{{2}}c_{{3}}S_{{3}}C_{{12}}+27\,{c_{{2}}}^{2}C_{{22}}C_{{12}}+12\,c_{{5}}C_{{12}}+18\,c_{{2}}c_{{5}}+9/2\,c_{{5}}S_{{3}}+9\,c_{{4}}C_{{13}}+3/2\,c_{{4}}S_{{4}}+6\,c_{{4}}C_{{22}} \nonumber \\
& & +15\,c_{{4}}{C_{{12}}}^{2} +45\,c_{{4}}{c_{{2}}}^{2}+33\,c_{{3}}c_{{4}}+81\,{c_{{3}}}^{2}C_{{12}}+ 33\,{c_{{3}}}^{2}S_{{3}}+90\,{c_{{3}}}^{2}c_{{2}}+48\,{c_{{2}}}^{3}c_{{3}}+3\,c_{{3}}C_{{23}}+3\,c_{{3}}C_{{14}} +48\,{c_{{2}}}^{3}C_{{22}} \nonumber \\
& & +18\,{c_{{2}}}^{2}C_{{23}}+36\,{c_{{2}}}^{3}C_{{13}}+69\,{c_{{2}}}^{3}{C_{{12}}}^{2}+12\,{c_{{2}}}^{4}S_{{3}}+6\,{c_{{2}}}^{2}C_{{14}}+54\,{c_{{2}}}^{4}C_{{12}}  +3\,{c_{{2}}}^{3}S_{{4}} \big ) {\kappa_{11}}^{2}{\kappa_{{2}}}^{3}  +{b_{{1}}}^{5} \big ( 6\,c_{{5}}C_{{12}}\nonumber \\
& & + 6\,c_{{4}}{C_{{12}}}^{2}+27\,c_{{2}}c_{{3}}C_{{22}}+13\,c_{{2}}c_{{5}}+c_{{6}}+96\,c_{{2}}c_{{4}}C_{{12}} +99\,{c_{{2}}}^{3}{C_{{12}}}^{2}+12\,{c_{{2}}}^{4}S_{{3}}+102\,{c_{{2}}}^{3}c_{{3}}+51\,{c_{{3}}}^{2}C_{{12}} \nonumber \\
& & +22\,c_{{3}}c_{{4}} +42\,{c_{{2}}}^{2}c_{{3}}S_{{3}}+360\,c_{{3}}{c_{{2}}}^{2}C_{{12}}+c_{{5}}S_{{3}}+105\,c_{{2}}c_{{3}}{C_{{12}}}^{2}+96\,{c_{{2}}}^{4}C_{{12}}+69\,{c_{{3}}}^{2}c_{{2}}+ 3\,{c_{{3}}}^{2}S_{{3}} +13\,c_{{2}}c_{{4}}S_{{3}}\nonumber \\
& & +78\,c_{{4}}{c_{{2}}}^{2}+24\,{c_{{2}}}^{3}C_{{22}}+30\,c_{{2}}c_{{3}}C_{{13}}+36\,{c_{{2}}}^{3}C_{{13}}+3\,c_{{4}}C_{{13}} \big ) {\kappa_{11}}^{3}{\kappa_{{2}}}^{2}+ {b_{{1}}}^{5} \big ( 6\,{c_{{2}}}^{5}+21\,c_{{2}}c_{{4}}C_{{12}}\nonumber \\
& & +126\,c_{{3}}{c_{{2}}}^{2}C_{{12}}+54\,{c_{{3}}}^{2}c_{{2}}+27\,{c_{{3}}}^{2}C_{{12}}+27\,c_{{4}}{c_{{2}}}^{2}+36\,{c_{{2}}}^{4}C_{{12}}+  6\,c_{{2}}c_{{5}}+48\,{c_{{2}}}^{3}c_{{3}}+9\,c_{{3}}c_{{4}} \big ) {\kappa_{11}}^{4}\kappa_{{2}}\nonumber \\
& & +12\,{b_{{1}}}^{5}{c_{{2}}}^{3}c_{{3}}{\kappa_{11}}^{5}+3\,{b_{{1}}}^{5}c_{{3}}c_{{4}}{\kappa_{11}}^{5}+9\,{b_{{1}}}^{5}{c_{{3}}}^{2}c_{{2}}{\kappa_{11}}^{5}+6\,{b_{{1}}}^{5}{c_{{2}}}^{2}c_{{4}}{\kappa_{11}}^{5}
\end {eqnarray}

\bsp

\label{lastpage}

\end{document}